\documentclass[journal,draftclsnofoot,onecolumn,12pt,oneside]{IEEEtran} 

\usepackage{cite}

\ifCLASSINFOpdf
  \usepackage[pdftex]{graphicx}
  \usepackage{epstopdf} 
\else
\fi
\usepackage{amsmath}
\usepackage{algorithm}
\usepackage{algorithmic}

  \usepackage[caption=false,font=footnotesize]{subfig}
\usepackage{url}

\usepackage{amsthm}

\newtheoremstyle{mydefinition}
{}
{}
{}
{0pt}
{\bfseries}
{.}
{ }
{\thmname{#1}\thmnumber{ #2}: \thmnote{#3}}

\theoremstyle{mydefinition}

\newtheoremstyle{myremark}
{}
{}
{}
{0pt}
{\bfseries}
{.}
{ }
{\thmname{#1}\thmnumber{ #2}: \thmnote{#3}}

\theoremstyle{theoremdd}

\theoremstyle{myremark}
\newcounter{rem}
\newtheorem{remark}[rem]{Remark}

\newcommand{\comment}[1]{{}}

\usepackage{amssymb}
\usepackage{xspace}
\usepackage{bm}
\usepackage{bbm}

\newcommand{\set}[1]{\ensuremath{\mathcal{#1}}\xspace} 
\newcommand{\mat}[1]{\ensuremath{\mathbf{#1}}\xspace} 
\renewcommand{\vec}[1]{\ensuremath{\mathbf{#1}}\xspace} 

\newcommand{\parens}[1]{\ensuremath{\left(#1\right)}\xspace}
\newcommand{\brackets}[1]{\ensuremath{\left[#1\right]}\xspace}
\newcommand{\braces}[1]{\ensuremath{\left\{#1\right\}}\xspace}
\newcommand{\bars}[1]{\ensuremath{\left\vert#1\right\vert}\xspace}
\newcommand{\doublebars}[1]{\ensuremath{\left\Vert#1\right\Vert}\xspace}

\renewcommand{\parens}[1]{{\left(#1\right)}\xspace}
\renewcommand{\brackets}[1]{{\left[#1\right]}\xspace}
\renewcommand{\braces}[1]{{\left\{#1\right\}}\xspace}
\renewcommand{\bars}[1]{{\left\vert#1\right\vert}\xspace}
\renewcommand{\doublebars}[1]{{\left\Vert#1\right\Vert}\xspace}

\newcommand{\complex}{\ensuremath{\mathbb{C}}\xspace}

\newcommand{\card}[1]{\bars{#1}}

\newcommand{\setcomplex}{\ensuremath{\complex}}

\newcommand{\setmatrix}[3]{\ensuremath{#1^{#2 \times #3}}\xspace}

\newcommand{\setmatrixcomplex}[2]{\setmatrix{\setcomplex}{#1}{#2}}

\newcommand{\ctrans}{\ensuremath{^{{*}}}\xspace}

\newcommand{\logten}[1]{\ensuremath{\mathrm{log}_{10}\parens{#1}}}

\newcommand{\pnorm}[2]{\ensuremath{\doublebars{#2}_{#1}}\xspace}

\newcommand{\normtwo}[1]{\pnorm{2}{#1}}

\newcommand{\normfro}[1]{\pnorm{\mathrm{F}}{#1}}

\newcommand{\distgauss}[2]{\ensuremath{\mathcal{N}\parens{#1,#2}}\xspace} 

\newcommand{\ind}[1]{\ensuremath{\mathbbm{1}\braces{#1}}\xspace}

\newcommand{\cdf}[2]{\ensuremath{\mathbb{F}_{#1}\parens{#2}}\xspace}

\newcommand{\maxop}[1]{\ensuremath{\mathrm{max}\parens{#1}}\xspace}

\newcommand{\minop}[1]{\ensuremath{\mathrm{min}\parens{#1}}\xspace}

\newcommand{\todB}[1]{\ensuremath{\brackets{#1}_{\mathrm{dB}}}}
\newcommand{\todBm}[1]{\ensuremath{\brackets{#1}_{\mathrm{dBm}}}}

\newcommand{\eirp}{\ensuremath{\mathsf{EIRP}}\xspace}

\newcommand{\powernoise}{\ensuremath{P_{\mathrm{noise}}}\xspace}

\newcommand{\powertx}{\ensuremath{P_{\mathrm{tx}}}\xspace}

\newcommand{\powersi}{\ensuremath{P_{\mathrm{SI}}}\xspace}

\newcommand{\snr}{\ensuremath{\mathsf{SNR}}\xspace}

\newcommand{\sinr}{\ensuremath{\mathsf{SINR}}\xspace}

\newcommand{\inr}{\ensuremath{\mathsf{INR}}\xspace}

\newcommand{\atx}[1]{\ensuremath{\vec{a}_{\mathrm{tx}}\parens{#1}}\xspace}
\newcommand{\arx}[1]{\ensuremath{\vec{a}_{\mathrm{rx}}\parens{#1}}\xspace}

\newcommand{\numrays}{\ensuremath{N_{\mathrm{rays}}}\xspace}
\newcommand{\numclust}{\ensuremath{N_{\mathrm{clust}}}\xspace}

\newcommand{\Nt}{\ensuremath{N_\mathrm{t}}\xspace} 
\newcommand{\Nr}{\ensuremath{N_\mathrm{r}}\xspace} 

\newcommand{\nbr}{\ensuremath{\parens{\Delta\theta,\Delta\phi}}\xspace}
\newcommand{\nbrv}{\ensuremath{\parens{\Deltavartheta,\Deltavarphi}}\xspace}
\newcommand{\nbrd}{\ensuremath{\parens{\deltatheta,\deltaphi}}\xspace}

\newcommand{\nbroneone}{\ensuremath{\parens{1^\circ,1^\circ}}\xspace}

\newcommand{\nbrtwotwo}{\ensuremath{\parens{2^\circ,2^\circ}}\xspace}
\newcommand{\nbrthreethree}{\ensuremath{\parens{3^\circ,3^\circ}}\xspace}

\newcommand{\thph}{\ensuremath{\parens{\theta,\phi}}\xspace}
\newcommand{\thphtx}{\ensuremath{\parens{\thetatx,\phitx}}\xspace}
\newcommand{\thphrx}{\ensuremath{\parens{\thetarx,\phirx}}\xspace}
\newcommand{\thphtxrx}{\ensuremath{\parens{\thetatx,\phitx,\thetarx,\phirx}}\xspace}
\newcommand{\thphtxi}{\ensuremath{\parens{\thetatx\idx{i},\phitx\idx{i}}}\xspace}
\newcommand{\thphrxj}{\ensuremath{\parens{\thetarx\idx{j},\phirx\idx{j}}}\xspace}

\newcommand{\idx}[1]{\ensuremath{^{\parens{#1}}}\xspace}

\newcommand{\txdirsetcb}{\ensuremath{\set{A}_{\mathrm{tx}}}\xspace}
\newcommand{\rxdirsetcb}{\ensuremath{\set{A}_{\mathrm{rx}}}\xspace}

\newcommand{\thetatx}{\ensuremath{\theta_{\mathrm{tx}}}\xspace}
\newcommand{\phitx}{\ensuremath{\phi_{\mathrm{tx}}}\xspace}

\newcommand{\thetarx}{\ensuremath{\theta_{\mathrm{rx}}}\xspace}
\newcommand{\phirx}{\ensuremath{\phi_{\mathrm{rx}}}\xspace}

\newcommand{\varthetatx}{\ensuremath{\vartheta_{\mathrm{tx}}}\xspace}
\newcommand{\varphitx}{\ensuremath{\varphi_{\mathrm{tx}}}\xspace}

\newcommand{\varthetarx}{\ensuremath{\vartheta_{\mathrm{rx}}}\xspace}
\newcommand{\varphirx}{\ensuremath{\varphi_{\mathrm{rx}}}\xspace}

\newcommand{\anglediff}[1]{\ensuremath{\measuredangle\parens{#1}}\xspace}

\newcommand{\Ntx}{\ensuremath{N_{\mathrm{tx}}}\xspace}
\newcommand{\Nrx}{\ensuremath{N_{\mathrm{rx}}}\xspace}

\newcommand{\labeltx}{\mathrm{tx}}
\newcommand{\labelrx}{\mathrm{rx}}

\newcommand{\deltatheta}{\ensuremath{\delta\theta}\xspace}
\newcommand{\deltaphi}{\ensuremath{\delta\phi}\xspace}

\newcommand{\Deltavartheta}{\ensuremath{\Delta\vartheta}\xspace}
\newcommand{\Deltavarphi}{\ensuremath{\Delta\varphi}\xspace}

\newcommand{\mHbar}{\bar{\mH}\xspace}

\newcommand{\mHest}{\bar{\mH}\xspace}

\newcommand{\mHestk}{\bar{\mH}_{k}\xspace}

\newcommand{\sAtx}{\sA_{\labeltx}\xspace}
\newcommand{\sArx}{\sA_{\labelrx}\xspace}

\newcommand{\sKtx}{\sK_{\labeltx}\xspace}
\newcommand{\sKrx}{\sK_{\labelrx}\xspace}

\newcommand{\sMreal}{\tilde{\sM}\xspace}

\newcommand{\sTtxi}{\sT_{\labeltx}\idx{i}\xspace}
\newcommand{\sTrxj}{\sT_{\labelrx}\idx{j}\xspace}

\newcommand{\thtx}{\ensuremath{\theta_{\mathrm{tx}}}\xspace}
\newcommand{\phtx}{\ensuremath{\phi_{\mathrm{tx}}}\xspace}
\newcommand{\thrx}{\ensuremath{\theta_{\mathrm{rx}}}\xspace}
\newcommand{\phrx}{\ensuremath{\phi_{\mathrm{rx}}}\xspace}
\newcommand{\thtxi}{\ensuremath{\thtx\idx{i}}\xspace}
\newcommand{\phtxi}{\ensuremath{\phtx\idx{i}}\xspace}
\newcommand{\thrxj}{\ensuremath{\thrx\idx{j}}\xspace}
\newcommand{\phrxj}{\ensuremath{\phrx\idx{j}}\xspace}

\newcommand{\inrreal}{\tilde{\inr}}

\newcommand{\inrijest}{\hat{\inr}_{ij}}

\newcommand{\setinr}{\sI}
\newcommand{\setinrij}{\sI_{ij}}

\newcommand{\inrmnreal}{\tilde{\inr}_{mn}}

\newcommand{\Gsq}{G^{2}\xspace}

\newcommand{\Gsqest}{\bar{G}^{2}\xspace}
\newcommand{\GsqestdB}{\todB{\Gsqest}\xspace}

\newcommand{\inrmax}{\inr_{\mathrm{max}}\xspace}
\newcommand{\inrmin}{\inr_{\mathrm{min}}\xspace}

\newcommand{\Gammaij}{\ensuremath{\Gamma_{ij}}\xspace}

\newcommand{\ksij}{\ensuremath{\kappa_{ij}}\xspace}

\newcommand{\muij}{\ensuremath{\mu_{ij}}\xspace}
\newcommand{\varij}{\ensuremath{\sigma_{ij}^2}\xspace}

\newcommand{\muest}{\ensuremath{\hat{\mu}}\xspace}
\newcommand{\varest}{\ensuremath{\hat{\sigma}^2}\xspace}

\newcommand{\muijest}{\ensuremath{\hat{\mu}_{ij}}\xspace}
\newcommand{\varijest}{\ensuremath{\hat{\sigma}_{ij}^2}\xspace}

\newcommand{\varijbar}{\ensuremath{\bar{\sigma}_{ij}^2}\xspace}

\newcommand{\varbar}{\ensuremath{\bar{\sigma}^2}\xspace}

\newcommand{\mumnest}{\ensuremath{\hat{\mu}_{mn}}\xspace}
\newcommand{\varmnest}{\ensuremath{\hat{\sigma}_{mn}^2}\xspace}

\newcommand{\distgaussij}{\ensuremath{\mathcal{N}_{ij}}\xspace}
\newcommand{\distgaussijest}{\ensuremath{\hat{\mathcal{N}}_{ij}}\xspace}

\newcommand{\meanop}[1]{\ensuremath{\mathrm{mean}\parens{#1}}\xspace}

\newcommand{\varop}[1]{\ensuremath{\mathrm{var}\parens{#1}}\xspace}

\def\vf{{\vec{f}}}

\def\vw{{\vec{w}}}

\def\mH{{\mat{H}}}

\def\sA{{\set{A}}}

\def\sI{{\set{I}}}

\def\sK{{\set{K}}}

\def\sM{{\set{M}}}

\def\sT{{\set{T}}}

\usepackage{xspace}
\usepackage[acronym,nogroupskip,nonumberlist,nopostdot]{glossaries}
\makeglossaries

\newcommand{\steer}{\textsc{Steer}\xspace}

\usepackage{xcolor}
\newcommand{\edit}[1]{\textcolor{black}{#1}}

\newacronym{snr}{SNR}{signal-to-noise ratio}
\newacronym{sinr}{SINR}{signal-to-interference-plus-noise ratio}
\newacronym{inr}{INR}{interference-to-noise ratio}
\newacronym{sir}{SIR}{signal-to-interference ratio}
\newacronym{sqr}{SQR}{signal-to-quantization-noise ratio}
\newacronym{sqnr}{SQNR}{signal-to-quantization-plus-noise ratio}
\newacronym{ian}{IAN}{interference as noise}
\newacronym{ber}{BER}{bit error rate}
\newacronym{pn}{PN}{pseudorandom noise}
\newacronym{bfsk}{BFSK}{binary frequency shift keying}
\newacronym{fh}{FH}{frequency-hopped}
\newacronym{fh-bfsk}{FH-BFSK}{frequency-hopped binary frequency shift keying}
\newacronym{crc}{CRC}{cyclic redundancy check}
\newacronym{isi}{ISI}{intersymbol interference}
\newacronym{dsss}{DSSS}{direct-sequence spread spectrum}
\newacronym{ofdm}{OFDM}{orthogonal frequency-division multiplexing}
\newacronym{ofdma}{OFDMA}{orthogonal frequency-division multiple access}
\newacronym{sdr}{SDR}{software-defined radio}
\newacronym{tx}{TX}{transmitter}
\newacronym{rx}{RX}{receiver}
\newacronym{fdd}{FDD}{frequency-division duplexing}
\newacronym{tdd}{TDD}{time-division duplexing}
\newacronym{fdma}{FDMA}{frequency-division multiple access}
\newacronym{tdma}{TDMA}{time-division multiple access}
\newacronym{sdma}{SDMA}{space-division multiple access}
\newacronym[plural=MPCs]{mpc}{MPC}{multipath component}
\newacronym{mui}{MUI}{multi-user interference}
\newacronym{lsb}{LSB}{least significant bit}

\newacronym{qam}{QAM}{quadrature amplitude modulation}
\newacronym{mqam}{MQAM}{M-ary quadrature amplitude modulation}

\newacronym{ls}{LS}{least-squares}
\newacronym{lms}{LMS}{least mean squares}
\newacronym{rls}{RLS}{recursive least-squares}
\newacronym{rzf}{RZF}{regularized zero-forcing}
\newacronym{mmse}{MMSE}{minimum mean square error}
\newacronym{lmmse}{LMMSE}{linear minimum mean square error}
\newacronym{mse}{MSE}{mean square error}
\newacronym{fft}{FFT}{fast Fourier transform}
\newacronym{dft}{DFT}{discrete Fourier transform}
\newacronym{dtft}{DTFT}{discrete-time Fourier transform}
\newacronym{ctft}{CTFT}{continuous-time Fourier transform}
\newacronym{ml}{ML}{machine learning}
\newacronym[plural=NNs]{nn}{NN}{neural network}
\newacronym[plural=RNNs]{rnn}{RNN}{recurrent neural network}
\newacronym[plural=ADCs]{adc}{ADC}{analog-to-digital converter}
\newacronym[plural=DACs]{dac}{DAC}{digital-to-analog converter}
\newacronym[plural=FPGAs]{fpga}{FPGA}{field-programmable gate array}
\newacronym{evm}{EVM}{error vector magnitude}
\newacronym{enob}{ENOB}{effective number of bits}
\newacronym{zf}{ZF}{zero-forcing}
\newacronym{rv}{r.v.}{random variable}
\newacronym{omp}{OMP}{orthogonal matching pursuit}
\newacronym{svd}{SVD}{singular value decomposition}

\newacronym{sdp}{SDP}{semidefinite programming}
\newacronym{psd}{PSD}{positive semidefinite}
\newacronym{nsd}{NSD}{negative semidefinite}

\newacronym{ks}{K-S}{Kolmogorov-Smirnov}

\newacronym{mad}{MAD}{median absolute deviation around the median}

\newacronym{agc}{AGC}{automatic gain control}
\newacronym{rf}{RF}{radio frequency}
\newacronym{if}{IF}{intermediate frequency}
\newacronym{los}{LOS}{line-of-sight}
\newacronym{nlos}{NLOS}{non-line-of-sight}
\newacronym{ple}{PLE}{path loss exponent}
\newacronym[plural=dB,firstplural=decibels (dB)]{db}{dB}{decibel}
\newacronym[plural=dBm,firstplural=decibel milliwatts (dBm)]{dbm}{dBm}{decibel milliwatts}
\newacronym{pa}{PA}{power amplifier}
\newacronym{lna}{LNA}{low noise amplifier}
\newacronym{vga}{VGA}{variable gain amplifier}
\newacronym{cw}{CW}{continuous wave}
\newacronym{papr}{PAPR}{peak-to-average power ratio}
\newacronym{usrp}{USRP}{Universal Software Radio Peripheral}
\newacronym{irr}{IRR}{image rejection ratio}
\newacronym{lo}{LO}{local oscillator}
\newacronym{vm}{VM}{vector modulator}
\newacronym{mmwave}{mmWave}{millimeter wave}
\newacronym{eirp}{EIRP}{effective isotropic radiated power}
\newacronym{rsrp}{RSRP}{reference signal received power}

\newacronym{csma}{CSMA}{carrier-sense multiple access}
\newacronym{csmaca}{CSMA/CA}{carrier-sense multiple access with collision avoidance}
\newacronym{csmacd}{CSMA/CD}{carrier-sense multiple access with collision detection}
\newacronym{mac}{MAC}{medium access control}
\newacronym{phy}{PHY}{physical layer}
\newacronym{4g}{4G}{fourth generation}
\newacronym{lte}{LTE}{Long-Term Evolution}
\newacronym{4glte}{4G LTE}{\gls{4g} \gls{lte}}
\newacronym{5g}{5G}{fifth generation}
\newacronym{nr}{NR}{New Radio}
\newacronym{5gnr}{5G NR}{5G New Radio}
\newacronym{ieee}{IEEE}{Institute of Electrical and Electronics Engineers}
\newacronym{wifi}{Wi-Fi}{IEEE 802.11}
\newacronym{lan}{LAN}{local area network}
\newacronym{wlan}{WLAN}{wireless local area network}
\newacronym[plural=BSs]{bs}{BS}{base station}
\newacronym[plural=SBSs]{sbs}{SBS}{small-cell base station}
\newacronym[plural=FD-SBSs]{fdsbs}{FD-SBS}{\gls{fd}-enabled \gls{sbs}}
\newacronym[plural=MBSs]{mbs}{MBS}{macrocell base station}
\newacronym[plural=UEs]{ue}{UE}{user equipment}
\newacronym{ul}{UL}{uplink}
\newacronym{dl}{DL}{downlink}
\newacronym{qos}{QoS}{Quality of Service}
\newacronym{fcc}{FCC}{Federal Communications Commission}
\newacronym{iab}{IAB}{integrated access and backhaul}
\newacronym{fab}{FAB}{fixed access and backhaul}
\newacronym{hetnet}{HetNet}{heterogeneous network}

\newacronym{siso}{SISO}{single-input single-output}
\newacronym{mimo}{MIMO}{multiple-input multiple-output}
\newacronym{sumimo}{SU-MIMO}{single-user \gls{mimo}}
\newacronym{mumimo}{MU-MIMO}{multi-user \gls{mimo}}
\newacronym{bf}{BF}{beamforming}
\newacronym{ca}{CA}{constant amplitude}
\newacronym{ula}{ULA}{uniform linear array}
\newacronym{upa}{UPA}{uniform planar array}
\newacronym[\glslongpluralkey={angles of arrival}]{aoa}{AoA}{angle of arrival}
\newacronym[\glslongpluralkey={angles of departure}]{aod}{AoD}{angle of departure}
\newacronym{dof}{DoF}{degrees of freedom}
\newacronym{csi}{CSI}{channel state information}
\newacronym{csit}{CSIT}{\gls{csi} at the transmitter}
\newacronym{csir}{CSIR}{\gls{csi} at the receiver}
\newacronym{cs}{CS}{compressed sensing}

\newacronym{fd}{FD}{in-band full-duplex}
\newacronym{hd}{HD}{half-duplex}
\newacronym{si}{SI}{self-interference}
\newacronym{sic}{SIC}{self-interference cancellation}
\newacronym{soi}{SoI}{signal of interest}
\newacronym{asic}{A-SIC}{analog \acrlong{sic}}
\newacronym{dsic}{D-SIC}{digital \gls{sic}}
\newacronym{star}{STAR}{simultaneous transmit and receive}
\newacronym{warp}{WARP}{Wireless Open-Access Research Platform}
\newacronym{bfc}{BFC}{beamforming cancellation}
\newacronym{ipi}{IPI}{inter-panel-interference}
\newacronym{ipic}{IPIC}{inter-panel-interference cancellation}

\newacronym{qcqp}{QCQP}{quadratically-constrained quadratic programming}
\newacronym{pdf}{PDF}{probability density function}
\newacronym{cdf}{CDF}{cumulative density function}
\newacronym{iid}{i.i.d.}{independently and identically distributed}

\newacronym{elf}{ELF}{extremely low frequency}
\newacronym{slf}{SLF}{super low frequency}
\newacronym{ulf}{ULF}{ultra low frequency}
\newacronym{vlf}{VLF}{very low frequency}
\newacronym{lf}{LF}{low frequency}
\newacronym{mf}{MF}{medium frequency}
\newacronym{hf}{HF}{high frequency}
\newacronym{vhf}{VHF}{very high frequency}
\newacronym{uhf}{UHF}{ultra high frequency}
\newacronym{shf}{SHF}{super high frequency}
\newacronym{ehf}{EHF}{extremely high frequency}
\newacronym{thf}{THF}{tremendously high frequency}

\newacronym{wncg}{WNCG}{Wireless Networking and Communications Group}
\newacronym{linc}{LINC}{Laboratory of Informatics, Networks, and Communications}
\newacronym{ut}{UT Austin}{The University of Texas at Austin}
\newacronym{uiuc}{UIUC}{University of Illinois at Urbana-Champaign}
\newacronym{usc}{USC}{University of Southern California}
\newacronym{mit}{MIT}{Massachusetts Institute of Technology}
\newacronym{berkeley}{UC Berkeley}{University of California, Berkeley}
\newacronym{osu}{OSU}{Ohio State University}

\newcommand{\upa}{\gls{upa}\xspace}
\newcommand{\upas}{\glspl{upa}\xspace}

\newcommand{\mmwave}{\gls{mmwave}\xspace}
\newcommand{\mimo}{\gls{mimo}\xspace}

\newcommand{\sic}{\acrlong{sic}\xspace}

\newcommand{\iab}{\gls{iab}\xspace}

\newcommand{\gcdf}{\gls{cdf}\xspace}
\newcommand{\gpcdf}{\glspl{cdf}\xspace}

\newcommand{\gsnr}{\gls{snr}\xspace}
\newcommand{\ginr}{\gls{inr}\xspace}
\newcommand{\gsinr}{\gls{sinr}\xspace}

\newcommand{\gks}{\gls{ks}\xspace}

\newcommand{\secref}[1]{Section~\ref{#1}}
\newcommand{\subsecref}[1]{Subsection~\ref{#1}}
\newcommand{\tabref}[1]{Table~\ref{#1}}
\newcommand{\figref}[1]{\figurename~\ref{#1}}
\newcommand{\algref}[1]{Algorithm~\ref{#1}}

\usepackage{mathtools}

\begin{document}
    
%
\title{Spatial and Statistical Modeling of Multi-Panel Millimeter Wave Self-Interference}

%
%
%


\author{%
    Ian~P.~Roberts,~%
    Aditya Chopra,~%
    Thomas Novlan,~\\%
    Sriram Vishwanath,~%
    and Jeffrey~G.~Andrews%
    \thanks{I.~P.~Roberts, S.~Vishwanath, and J.~G.~Andrews are with 6G@UT in the Wireless Networking and Communications Group at the University of Texas at Austin. A.~Chopra was with the Advanced Wireless Technologies Group at AT\&T Labs during this work; he is now with Project Kuiper at Amazon. T.~Novlan is with the Advanced Wireless Technologies Group at AT\&T Labs.}%
    \thanks{\edit{Related code is available at: {https://ianproberts.com/simodel}.}}
}

\maketitle





\begin{abstract}
Characterizing self-interference is essential to the design and evaluation of in-band full-duplex communication systems.
Until now, little has been understood about this coupling in full-duplex systems operating at \mmwave frequencies, and it has been shown that the highly-idealized models proposed for such do not align with practice.
This work presents the first spatial and statistical model of \mmwave self-interference backed by measurements, enabling engineers to draw realizations that exhibit the large-scale and small-scale spatial characteristics observed in our nearly 6.5 million measurements taken at 28 GHz.
Core to our model is its use of system and model parameters having real-world meaning, which facilitates its extension to systems beyond our own phased array platform through proper parameterization.
We demonstrate this by collecting nearly 13 million additional measurements to show that our model can generalize to two other system configurations. 
We assess our model by comparing it against actual measurements to confirm its ability to align spatially and in distribution with real-world self-interference.
In addition, using both measurements and our model of self-interference, we evaluate an existing beamforming-based full-duplex \mmwave solution to illustrate that our model can be reliably used to design new solutions and validate the performance improvements they may offer. 
\end{abstract}




\glsresetall


\section{Introduction} \label{sec:introduction}

Full-duplex \mmwave communication systems have drawn increased attention recently due to their potential enhancements and applications in next-generation wireless networks \cite{xia_2017,roberts_wcm,gupta_fdiab,3GPP_IAB_2,liu_beamforming_2016}.
The prospects of full-duplex in \mmwave systems are particularly exciting thanks to ongoing efforts demonstrating the cancellation of self-interference purely through beamforming \cite{liu_beamforming_2016,satyanarayana_hybrid_2019,cai_robust_2019,lopez_analog_2022,koc_ojcoms_2021,da_silva_2020,roberts_bflrdr,roberts_lonestar}. 
By strategically steering or shaping its transmit and receive beams, it has been shown that a full-duplex \mmwave system can reduce self-interference to levels near or below the noise floor, facilitating simultaneous transmission and reception across the same frequency spectrum---without requiring analog or digital \sic. 
The success of such beamforming-based approaches, however, has almost exclusively been validated through simulation using highly-idealized self-interference channel models that have not been verified through measurement.
In fact, until now, very little has been understood about the spatial composition of \mmwave self-interference in real-world systems, raising questions regarding the efficacy of these proposed solutions and of full-duplex \mmwave systems altogether.
A measurement-backed model of \mmwave self-interference would facilitate accurate design and evaluation of practical full-duplex \mmwave systems.



\subsection{Existing Models and Measurements} 

The most popular way to model self-interference in \mmwave systems thus far has been based on near-field propagation via the spherical-wave \mimo channel model \cite{spherical_2005}.
Albeit sensible, this channel model is extremely idealized and ignores a multitude of real-world factors including non-isotropic antenna array elements, array enclosures, mounting infrastructure, and a variety of other nonidealities.
Furthermore, this channel model is deterministic for a given relative transmit and receive array geometry, which prevents engineers from capturing artifacts unique to individual full-duplex \mmwave systems.
This spherical-wave model has often been used in conjunction with a ray-based model \cite{satyanarayana_hybrid_2019,li_2014} to capture reflections off the environment in addition to the direct coupling between the transmit and receive arrays.
Through simulation, these models of self-interference have been used widely to develop and validate beamforming-based solutions for full-duplex \mmwave systems. 
However, these approaches to model self-interference have not been verified through measurement, meaning it is unclear if proposed solutions evaluated under such models would actually see success in practice. 

Nevertheless, there have been efforts to measure and characterize \mmwave self-interference.
In \cite{rajagopal_2014}, for instance, measurements of 28 GHz self-interference power were reported for 8$\times$8 \upas in indoor and outdoor environments.
These measurements are certainly valuable but offer limited insights on the spatial and statistical characteristics of self-interference.
The works of \cite{lee_2015,yang_2016,he_2017,haneda_2018} also collected measurements of \mmwave self-interference using horn and lens antennas, which shed light on environmental factors and the power delay profile of self-interference.
However, the measurements in \cite{rajagopal_2014,lee_2015,yang_2016,he_2017,haneda_2018} did not provide extensive characterization or modeling of self-interference (not a small task), limiting their usefulness in the development and evaluation of full-duplex \mmwave systems---especially those employing phased arrays and relying on beamforming to cancel self-interference.

Thus far, the most extensive measurement campaign and characterization of \mmwave self-interference is our recent work \cite{roberts_att_angular}.
Therein, we collected nearly 6.5 million measurements of 28 GHz self-interference using 16$\times$16 \upas, whose analysis revealed high-level spatial trends and noteworthy small-scale variability. 
We provided a thorough statistical characterization of measured self-interference to facilitate the design and evaluation of full-duplex \mmwave systems. 
In \cite{roberts_att_angular}, however, we made no attempt to model the \textit{spatial} composition of self-interference but did show that the spherical-wave channel model \cite{spherical_2005} does not align with our measurements---motivating the need for a new model fit for real-world full-duplex \mmwave systems.

\subsection{Contributions}
In this paper, we introduce the first measurement-backed model of self-interference in multi-panel\footnote{We use the term ``multi-panel'' to refer to separate transmit and receive arrays.} full-duplex \mmwave systems.
Using measurements of 28 GHz self-interference collected in our campaign \cite{roberts_att_angular}, we construct a novel stochastic model that aligns both \textit{statistically} and \textit{spatially} with real-world self-interference.
Our model is relatively simple, has theoretical foundations, and is built on system and model parameters with physical meaning.
This gives our model the potential to generalize to systems beyond our own \edit{28 GHz} phased array platform via unique parameterization.
We demonstrate this by fitting our model to nearly 13 million additional measurements collected from two other system configurations. 
To construct our model, we uncovered a coarse geometric approximation of the self-interference channel from within our measurements, which suggests that the dominant coupling between the transmit and receive arrays manifests as clusters of rays in a far-field manner, rather than in a near-field, spherical-wave fashion.
This is a novel finding that can steer future work aiming to model self-interference \mimo channels in full-duplex \mmwave systems.



By comparing statistical realizations of self-interference from our model to actual measurements, we show that our model exhibits the large-scale and small-scale spatial characteristics of self-interference observed in real full-duplex \mmwave systems.
In fact, to further confirm this, we evaluate an existing full-duplex solution \cite{roberts_steer} using our model and repeat this evaluation using actual measurements.
The coinciding results of these two evaluations indicates that our model can indeed serve as a useful tool in the design and validation of full-duplex \mmwave systems by giving engineers the ability to draw accurate realizations of self-interference and conduct statistical analyses.
\edit{Simultaneously, we also illustrate that the commonly-used spherical-wave model \cite{spherical_2005} neither spatially nor statistically aligns with our measurements of self-interference.}

\edit{%
The present work and our prior work \cite{roberts_att_angular} can both be used to draw statistical realizations of \mmwave self-interference, but the two differ notably.
With \cite{roberts_att_angular}, statistical realizations must simply be drawn from the global distribution of our measurements and therefore lack any spatial consistency.
With the model presented herein, realizations can be drawn that are both statistically and spatially consistent with our measurements, courtesy of the spatial model of self-interference that we have uncovered.
This spatial modeling sheds light on the underlying nature of self-interference, with the consequent spatial consistency facilitating more accurate design and evaluation of full-duplex \mmwave systems.
}

\textit{Notation:} 
We use $\distgauss{\mu}{\sigma^2}$ to denote the normal distribution having mean $\mu$ and variance $\sigma^2$.
We use $\theta$ and $\phi$ as directions in azimuth and elevation, respectively.
We use $i$ and $j$ for transmit and receive indexing, respectively.
We use hat $\hat{a}$ to denote the estimate of some ground truth $a$. 
We use bar $\bar{A}$ to denote a model parameter having connections to some theoretical system parameter $A$.
All symbols have linear units by default; we use $\todB{A} = 10\cdot\logten{A}$ to denote $A$ in units of decibels (dB) and $\todBm{B}$ to denote $B$ in units of decibel-milliwatts (dBm).

\comment{
---

In this work, we present the first measurement-backed stochastic model of \mmwave self-interference. 
The construction of our model is driven by nearly 6.5 million measurements of 28 GHz self-interference power using 16$\times$16-element phased arrays.
We leverage large-scale spatial trends and small-scale spatial variability observed in our measurements to construct a novel stochastic model of self-interference.
Engineers can use our stochastic model to draw realizations of self-interference that are spatially and statistically consistent with measurements taken with an actual multi-panel full-duplex \mmwave system---making our model the first of its kind in this regard.

Our model of self-interference is parameterized by various system and model variables that allow it to potentially translate to other systems beyond our phased array platform with appropriate parameterization.
Core to our model is a coarse approximation of the self-interference channel matrix inferred from our measurements, which suggests that the dominant source of self-interference propagates along multiple clusters of rays from the transmit array to the receive array. 
This is a novel finding that can inspire future work aiming to model self-interference \mimo channels in full-duplex \mmwave systems.

Upon evaluation, our model proves that it is capable of producing realizations of self-interference that are statistically and spatially aligned with actual measurements.
We conduct a performance evaluation of an existing full-duplex solution \cite{roberts_steer} using our model and compare the results to an evaluation using actual measurements.
The coinciding results of these evaluations confirms that our model can serve as a useful tool in the design and evaluation of practical full-duplex \mmwave solutions.
}

\section{Measurements of mmWave Self-Interference} \label{sec:setup}

\begin{figure*}
    \centering
    \subfloat[28 GHz phased array platform.]{\includegraphics[width=0.35\linewidth,height=\textheight,keepaspectratio]{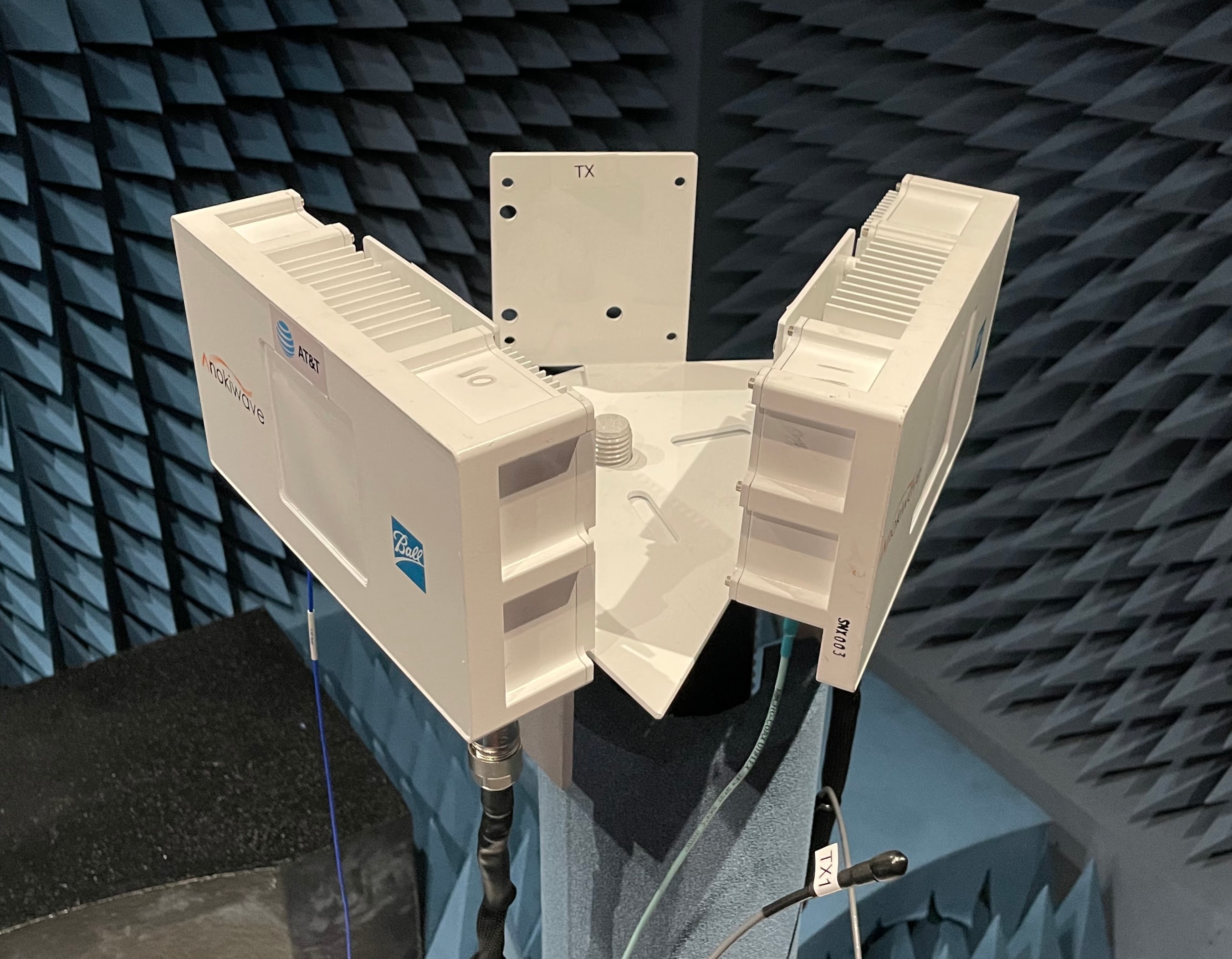}
        \label{fig:setup}}
    \qquad
    \subfloat[Multi-panel full-duplex base station.]{\includegraphics[width=0.55\linewidth,height=\textheight,keepaspectratio]{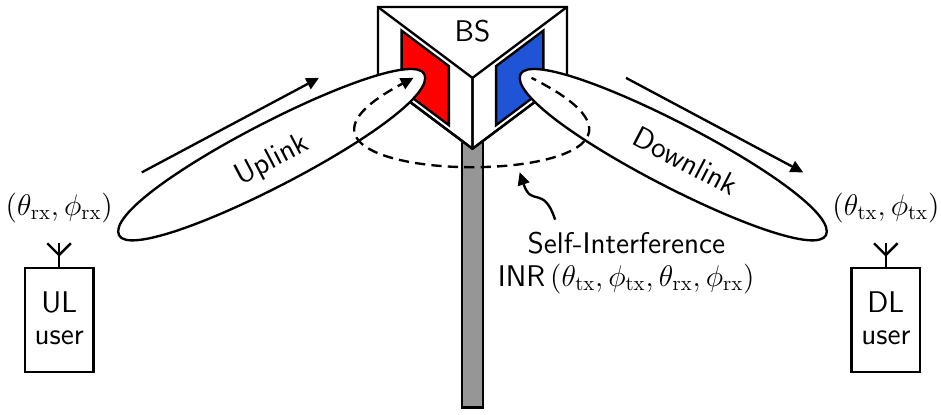}
        \label{fig:system}}
    \caption{(a) Our 28 GHz measurement platform in an anechoic chamber. The transmit array (right) and receive (left) array are identical 16$\times$16 planar arrays \cite{anokiwave}. (b) A full-duplex \mmwave base station transmits downlink with one array while receiving uplink with another array in-band. In doing so, self-interference couples between the arrays, the degree of which depends on the transmit and receive beams and the self-interference channel.}
    \label{fig:setup-system}
\end{figure*}

The work herein is based on data collected in our recent measurement campaign of 28~GHz self-interference \cite{roberts_att_angular}.
In this section, we will summarize our measurement platform and methodology, highlighting key and necessary information. 
We collected measurements of self-interference incurred by a multi-panel full-duplex \mmwave system employing two phased arrays mounted on separate sides of a sectorized triangular platform, as shown in \figref{fig:setup-system}.
One phased array acted as a transmitter while the other served as a receiver, with their centers separated by 30 cm.
The transmitter and receiver are identical Anokiwave AWA-0134 28 GHz phased arrays \cite{anokiwave}, each of which contains a 16$\times$16 half-wavelength \gls{upa}.
This sectorized configuration with industry-grade phased arrays aligns with realistic multi-panel full-duplex \mmwave deployments, such as for \iab as proposed in 3GPP \cite{3GPP_IAB_2}.
Measurements were conducted in an anechoic chamber, free from significant reflectors, to inspect the direct coupling between the arrays; investigating self-interference in real-world environments is necessary future work. 

\begin{figure}
    \centering
    \includegraphics[width=\linewidth,height=0.25\textheight,keepaspectratio]{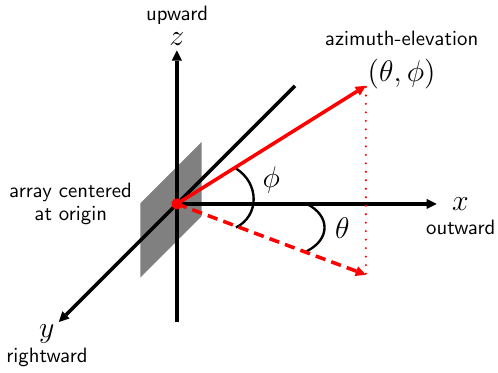}
    \caption{\edit{The coordinate system observed by each phased array situated at its respective origin, facing toward the positive $x$ axis. An azimuth-elevation of $(0^\circ,0^\circ)$ corresponds to steering outward (broadside) along the $x$ axis. An increase in azimuth is rightward, toward the $y$ axis. An increase in elevation is upward, toward the $z$ axis.}}
    \label{fig:geometry}
\end{figure}

Each phased array can be electronically steered toward some azimuth and elevation via digitally-controlled analog beamforming weights.
With 256 antenna elements, our phased arrays can produce highly directional beams, enabling us to inspect self-interference with fine spatial granularity.
To steer our transmit beam or receive beam toward some azimuth-elevation $\thph$, we use conjugate beamforming \cite{heath_lozano}, where the transmit and receive beamforming vectors are 
\begin{align}
\vf\thph &= \atx{\theta,\phi}, \qquad
\vw\thph = \arx{\theta,\phi}. \label{eq:cbf-f-w}
\end{align}
Here, $\atx{\theta,\phi}$ and $\arx{\theta,\phi}$ are the array response vectors of our transmit array and receive array in some direction $\thph$.
These array response vectors are solely a characteristic of the array geometries and carrier frequency and can be computed in a closed-form \cite{heath_lozano,balanis}.
\edit{We employ the 3-D geometry shown in \figref{fig:geometry}, where the steering direction of each phased array is relative to its own local coordinate system.}
Our array response vectors and beamforming vectors are normalized as
\begin{align}
\normtwo{\vf\thph}^2 &= \normtwo{\atx{\theta,\phi}}^2 = \Nt, \qquad
\normtwo{\vw\thph}^2 = \normtwo{\arx{\theta,\phi}}^2 = \Nr, \label{eq:beams-f-w}
\end{align}
where $\Nt = \Nr = 256$ is the number of transmit and receive elements in our particular arrays.


Assuming a linear setting, when transmitting toward some direction $\thphtx$ and receiving toward $\thphrx$, the coupled self-interference power at the receiver output can be expressed as
\begin{align}
\powersi\parens{\thetatx,\phitx,\thetarx,\phirx} = \eirp \cdot \Gsq \cdot \underbrace{\bars{\vw\parens{\thetarx,\phirx}\ctrans \mH \vf\parens{\thetatx,\phitx}}^2}_{\mathsf{coupling~factor}}, \label{eq:power-si} 
\end{align}
where $\mH \in \setmatrixcomplex{\Nr}{\Nt}$ is the self-interference \mimo channel manifesting between the transmit and receive arrays and $\eirp = 60$ dBm is the \gls{eirp} of our transmitting phased array.
Here, the scalar $G^2$ captures the inherent path loss between the arrays, along with amplification and losses internal to the transmit and receive array modules. 
Rather than raw power, it is perhaps more useful to use an \ginr to characterize self-interference.
As such, we normalize received self-interference power to the noise floor as
\begin{align}
\inr\parens{\thetatx,\phitx,\thetarx,\phirx} = \frac{\powersi\parens{\thetatx,\phitx,\thetarx,\phirx}}{\powernoise}, \label{eq:power-si-inr}
\end{align}
where $\powernoise = -68$ dBm is the integrated noise power at our receive array output over our 100~MHz measurement bandwidth.
We used Zadoff-Chu sequences and correlation-based processing to reliably measure received self-interference power well below the noise floor.
Using high-fidelity test equipment and stepped attenuators, we calibrated our measurement capability and verified that it had low error (typically less than 1 dB) broadly across received power levels.

With measures of \ginr, full-duplex system performance can be quantified by computing the received \gsinr as
\begin{align}
\sinr\parens{\thetatx,\phitx,\thetarx,\phirx} 
&= \frac{\snr}{1 + \inr\parens{\thetatx,\phitx,\thetarx,\phirx}},
\end{align}
where $\snr$ is the \gsnr of a desired receive signal at the output of the receive array, which would practically depend on $\thphrx$.
To solely understand the degree of self-interference at the receive array output, this work is concerned with measuring and modeling $\inr$ when no further cancellation is employed (i.e., no analog or digital \sic).
\ginr alone is useful in indicating if a full-duplex system is noise-limited ($\inr \ll 0$ dB) or self-interference-limited ($\inr \gg 0$ dB).
Such systems desire a low \ginr, say $\inr < 0$ dB in most cases, to ensure self-interference does not erode full-duplexing gains.

To collect \ginr measurements over a broad and dense spatial profile, we swept the transmit beam and receive beam across a number of directions in azimuth and elevation.
Prior to measurement, we specified sets of $\Ntx$ transmit directions and $\Nrx$ receive directions
\begin{align}
\txdirsetcb &= \braces{\parens{\thetatx\idx{i},\phitx\idx{i}}}_{i=1}^{\Ntx}, \qquad
\rxdirsetcb = \braces{\parens{\thetarx\idx{j},\phirx\idx{j}}}_{j=1}^{\Nrx}
\end{align}
over which the phased arrays were electronically swept, measuring self-interference for each transmit-receive combination.
In this work, we densely swept the transmit and receive beams from $-60^\circ$ to $60^\circ$ in azimuth and from $-10^\circ$ to $10^\circ$ in elevation, both in $1^\circ$ steps.
\begin{align}
\txdirsetcb = \rxdirsetcb 
&= \underbrace{\braces{-60^\circ,-59^\circ,\dots,60^\circ}}_{\mathsf{azimuth}} \times \underbrace{\braces{-10^\circ,-9^\circ,\dots,10^\circ}}_{\mathsf{elevation}} \label{eq:spatial-profile} \\
&= \braces{\parens{-60^\circ,-10^\circ},\dots,\parens{-60^\circ,10^\circ},\dots,\parens{60^\circ,-10^\circ},\dots,\parens{60^\circ,10^\circ}}
\end{align}
This amounts to $\Ntx = \Nrx = 121 \times 21 = 2541$ steering directions for a total of $\Ntx \times \Nrx \approx 6.5$ million \ginr measurements.
The set of nearly 6.5 million \ginr measurements we denote as
\begin{align}
\setinr = \braces{\inr\thphtxrx: \thphtx \in \sAtx, \thphrx \in \sArx}.
\end{align}
From this set of measurements, we spatially and statistically model self-interference in the sections that follow.
For a summary of these collected measurements, please see \cite{roberts_att_angular}.

\section{Motivation and Roadmap} \label{sec:motivation}




The goal of this work is to produce a stochastic model of \mmwave self-interference that is statistically and spatially well-aligned with our collected measurements.
Such a model would allow engineers to draw countless realizations of self-interference and facilitate accurate simulation, development, and evaluation of full-duplex \mmwave systems.
Even if one were to assume the same setup as ours, it is unrealistic to assume our measurements could be directly extended to other systems due to beam-steering and positioning differences, among numerous other practical artifacts.
As such, being able to draw \textit{realizations} of self-interference---as opposed to estimating deterministic values---will be more generalizable and useful to the research community. 

\begin{figure*}
    \centering
    \includegraphics[width=\linewidth,height=0.23\textheight,keepaspectratio]{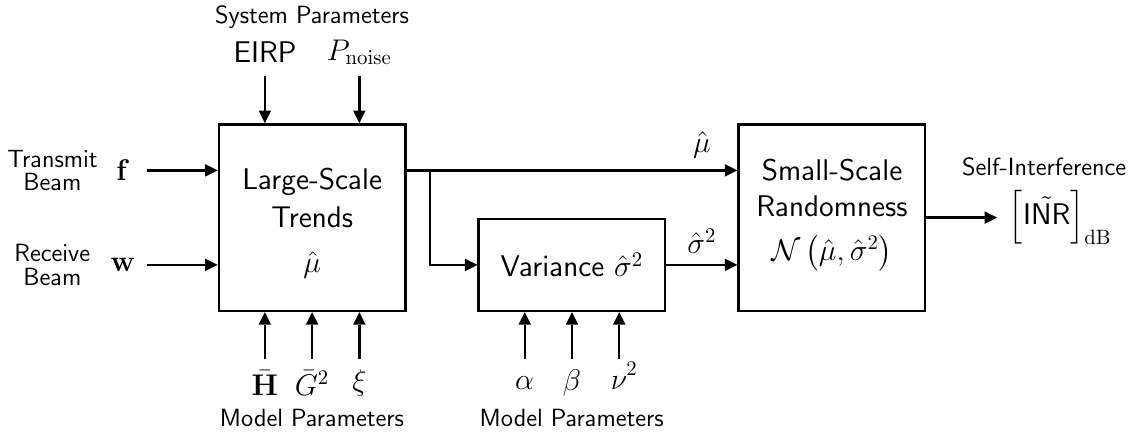}
    \caption{A block diagram of our statistical model of self-interference. Based on system parameters, model parameters, and transmit and receive beams, our model produces realizations of self-interference that are statistically and spatially aligned with actual measurements.}
    \label{fig:model}
\end{figure*}


As we will elaborate on shortly, our statistical model of self-interference is based on two characteristics observed in our measurements:
\begin{enumerate}
    \item On a large scale (at a high level), there is a connection between the steering directions of the transmit and receive beams and the degree of self-interference incurred.
    Broadly speaking, some transmit and receive directions tend to incur high self-interference while others tend to incur low self-interference.
    \item On a small scale (within small spatial neighborhoods), the system incurs seemingly random amounts of self-interference.
    Slightly shifting the transmit and receive steering directions can dramatically alter the degree of self-interference coupled.
\end{enumerate}
We leverage these large-scale and small-scale characteristics to construct a stochastic model of self-interference that both statistically and spatially aligns with our measurements.

A block diagram summarizing our model is shown in \figref{fig:model}, which will become clear as we proceed in its presentation.
For particular transmit and receive beams $\vf$ and $\vw$, a mean parameter $\muest$ is estimated, which dictates the location of the distribution from which self-interference is drawn.
The variance of this distribution $\varest$ is dictated by $\muest$ and other model parameters.
This approach allows our model to capture the large-scale spatial trends in self-interference along with the small-scale variability observed over small spatial neighborhoods.
With appropriate parameterization, our model has the potential to be extended to other systems and environments beyond our own, which we demonstrate in \secref{sec:remarks} by fitting our model to measurements collected with two other system configurations. 

\section{Log-Normally-Distributed Self-Interference\\Over Small Spatial Neighborhoods} \label{sec:lognormal}




\begin{figure}
    \centering
    \includegraphics[width=\linewidth,height=0.2\textheight,keepaspectratio]{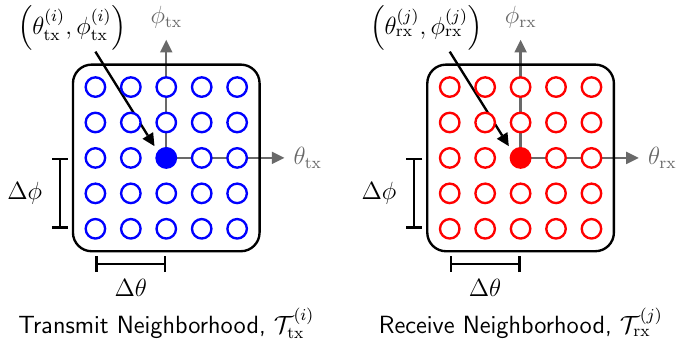}
    \caption{\edit{Spatial neighborhoods of size $\nbr$ surrounding the $i$-th transmit direction and $j$-th receive direction.}}
    \label{fig:nbr}
\end{figure}

In pursuit of a model of self-interference that is {stochastic}, we make use of a small-scale phenomenon present in our measurements: seemingly random amounts of self-interference are observed over small spatial neighborhoods.
To quantify this small-scale randomness, we introduce the concept of a \textit{spatial neighborhood}, \edit{illustrated in \figref{fig:nbr}}.
We begin by defining $\anglediff{\theta_1,\theta_2}$ as the absolute difference between two angles $\theta_1,\theta_2$ (in degrees), written as
\begin{align}
\anglediff{\theta_1,\theta_2} =
\begin{cases}
\zeta, & \zeta \leq 180^\circ \\
360^\circ - \zeta, & \zeta > 180^\circ \\
\end{cases}
\end{align}
where $\zeta = \bars{\theta_1 - \theta_2} \ \mathrm{mod} \ 360^\circ$ and $\mathrm{mod}$ is the modulo operator.
Let $\sTtxi\nbr$ be the spatial neighborhood of size \nbr in azimuth-elevation surrounding the $i$-th measured transmit direction $\thphtxi \in \sAtx$, defined as follows, and let $\sTrxj\nbr$ analogously be the spatial neighborhood surrounding the $j$-th measured receive direction $\thphrxj \in \sArx$.
\begin{align}
\sTtxi\nbr &= \braces{\thph \in \sAtx : \anglediff{\theta,\thtxi} \leq \Delta\theta, \anglediff{\phi,\phtxi} \leq \Delta\phi} \\
\sTrxj\nbr &= \braces{\thph \in \sArx : \anglediff{\theta,\thrxj} \leq \Delta\theta, \anglediff{\phi,\phrxj} \leq \Delta\phi}
\end{align}
In other words, $\sTtxi\nbr$ is the set of measured transmit directions within $\Delta\theta$ in azimuth and within $\Delta\phi$ in elevation of the $i$-th transmit direction $\thphtxi$. 
Using these, let $\setinrij\nbr$ be the set of measured \ginr values over transmit and receive spatial neighborhoods surrounding the $(i,j)$-th beam pair.
\begin{align}
\setinrij\nbr = \braces{\inr\thphtxrx : \thphtx \in \sTtxi\nbr, \thphrx \in \sTrxj\nbr}
\end{align}

We observed that \ginr values over spatial neighborhoods of size $\nbr = \nbrtwotwo$ were approximately log-normally distributed throughout our measurements.
Put simply, we found that 
\begin{align}
\todB{\inr \in \setinrij\nbrtwotwo} \overset{\mathrm{approx.}}{\sim} \distgauss{\muij}{\varij} \ \forall \ i,j,
\end{align}
where $\muij$ and $\varij$ are the mean and variance of the normal distribution fitted to $\todB{\setinrij\nbrtwotwo}$.
Note that $\muij$ and $\varij$ are specific to the neighborhood surrounding a particular transmit-receive beam pair $(i,j)$, meaning there are nearly 6.5 million fitted $\parens{\muij,\varij}$.
Also, note that $\muij$ implicitly has units of decibels.
Three instances of $\setinrij\nbrtwotwo$ depicting this log-normal nature can be observed in \figref{fig:lognormal-a}, each of which has a unique fitted mean $\muij$ and variance $\varij$.
\edit{Each of these three \gpcdf in \figref{fig:lognormal-a} illustrates that slightly shifting the transmit and receive beams leads to wide variability in self-interference.}

\subsection{Confirming Log-Normal Nature Across All Measurements}
To quantitatively evaluate this log-normal nature over our entire set of measurements, we computed the \gks statistic of the fit for each $(i,j)$ as follows.
Henceforth, under the assumption of a $\nbr = \nbrtwotwo$ spatial neighborhood, we employ the shorthand
\begin{align}
\setinrij \triangleq \todB{\setinrij\nbrtwotwo}.
\end{align}
Let $\cdf{\setinrij}{X}$ be the empirical \gcdf of $\setinrij$---that is, the fraction of elements in $\setinrij$ less than or equal to $X$, defined as
\begin{align}
\cdf{\setinrij}{X} = \frac{1}{\card{\setinrij}} \cdot \sum_{\inr \in \setinrij} \ind{\inr \leq X},
\end{align}
where $\ind{\cdot}$ is the indicator function.
Similarly, let $\cdf{\distgaussij}{X}$ be the \gcdf of the normal distribution $\distgaussij \triangleq \distgauss{\muij}{\varij}$ evaluated at $X$.
For each $(i,j)$, the \gls{ks} statistic $\ksij \in \brackets{0,1}$ can be written as
\begin{align}
\ksij &= \max_{X} \ \bars{\cdf{\setinrij}{X} - \cdf{\distgaussij}{X}},
\end{align}
which describes the maximum absolute difference in cumulative density between two distributions and therefore serves as a measure of similarity between the distributions.
A $\ksij \leq 0.1$ indicates that self-interference over the $(i,j)$-th neighborhood closely follows a log-normal distribution, whereas higher $\ksij$ indicates poorer fitting.

\begin{figure*}
    \centering
    \subfloat[Exemplar log-normally-distributed neighborhoods.]{\includegraphics[width=0.475\linewidth,height=0.27\textheight,keepaspectratio]{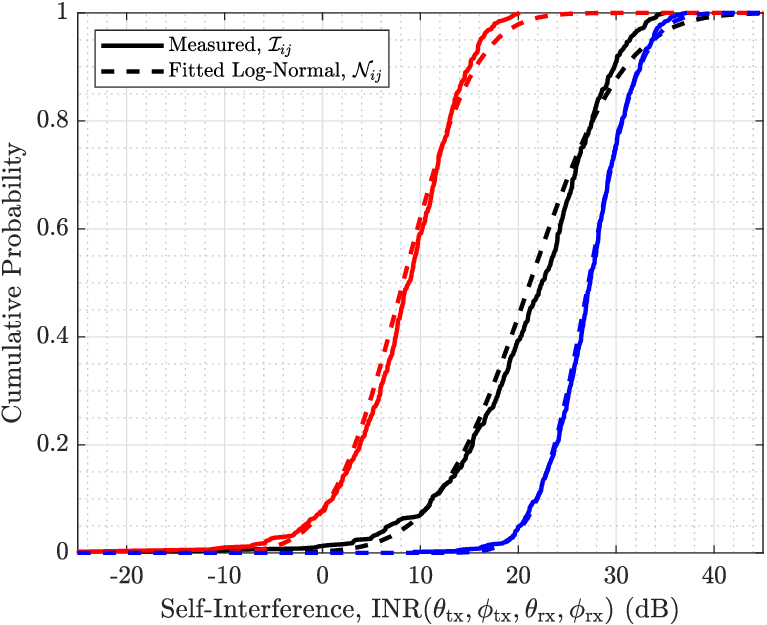}
        \label{fig:lognormal-a}}
    \quad
    \subfloat[K-S statistic $\ksij$ for various $\nbr$.]{\includegraphics[width=0.475\linewidth,height=0.27\textheight,keepaspectratio]{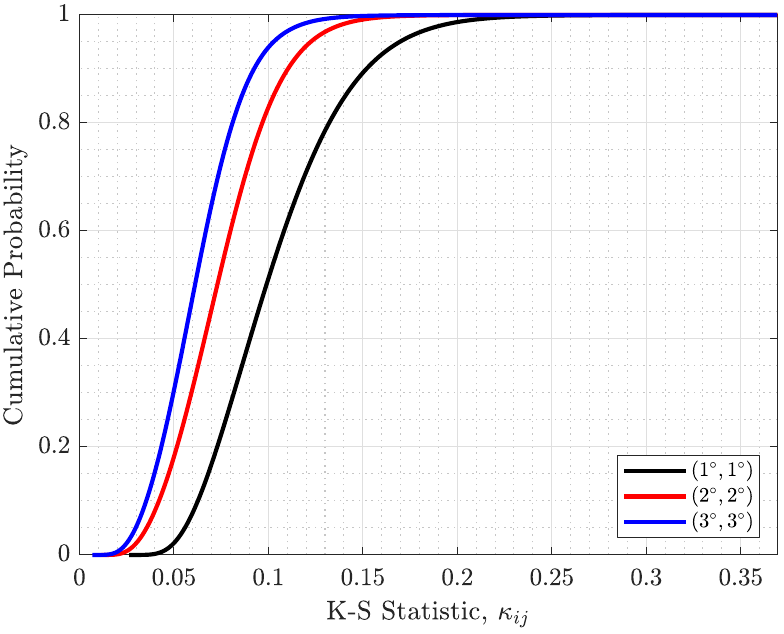}
        \label{fig:lognormal-b}}
    \caption{(a) Three spatial neighborhoods whose measured self-interference follows a log-normal distribution, each with a unique fitted mean and variance. (b) The \gpcdf of the \gls{ks} statistic $\ksij$ across all nearly 6.5 million beam pairs $(i,j)$ for various neighborhood sizes $\nbr$. The vast majority of measured beam pairs exhibit levels of self-interference that are log-normally distributed across the surrounding $\nbrtwotwo$ spatial neighborhood.}
    \label{fig:lognormal}
\end{figure*}

In \figref{fig:lognormal-b}, we plot the \gcdf of ${\ksij}$ over the nearly 6.5 million neighborhoods (i.e., for each $(i,j)$).
We have plotted this for a neighborhood size of $\nbrtwotwo$, as well as $\nbroneone$ and $\nbrthreethree$; for now, consider $\nbrtwotwo$.
We see that nearly $50$\% are within $\kappa_{ij} \leq 0.07$, over $80$\% are within $\kappa_{ij} \leq 0.1$, and nearly $95$\% are within $\kappa_{ij} \leq 0.12$.
This suggests that the vast large majority of neighborhoods are well approximated by their fitted log-normal distributions.
There exists a noteworthy upper tail with less than $1$\% of neighborhoods yielding $\kappa_{ij} \geq 0.15$, which indicates that select neighborhoods do not follow a log-normal distribution as closely as the overwhelming majority.
Nonetheless, the log-normal nature of local neighborhoods is the foundation for the statistical model of self-interference we present henceforth, and we acknowledge that it may limit how well our model can capture the outliers of our measurements. 

Compared to $\nbroneone$, a neighborhood size of $\nbrtwotwo$ more often closely follows a log-normal distribution, a notable improvement of about $0.03$ in $\kappa_{ij}$ in distribution.
A neighborhood size of $\nbrthreethree$ indeed offers improvements over $\nbrtwotwo$ in terms of log-normal fit, but these are marginal, making it more desirable to use a smaller neighborhood to allow our model to capture finer large-scale spatial variability.
This motivates our use of a neighborhood size of $\nbrtwotwo$ henceforth with the understanding that others may be suitable as well.
\edit{In addition, distributions other than log-normal could also used to capture this small-scale variability; we chose a log-normal because it fit well broadly across our measurements and because its statistical parameters can be captured succinctly with our proposed model, as we will see.}

\subsection{What Causes This Log-Normal Nature?}
\edit{%
It is difficult to precisely describe the roots of this log-normal nature, but we can make educated speculations based on an understanding of our measurement system and wireless fundamentals. 
First of all, self-interference is presumably coupled behind the arrays by their side lobes and back lobes, and when slightly shifting the steering directions of the main lobes, these side and back lobes fluctuate and the location of nulls move---this can lead to destructive combining when nulls manifest in directions of strong interference.
Moreover, in theory, arrays exhibit idealized beam patterns with well-defined nulls, which would make self-interference more straightforwardly understood.
In practice, however, beam shapes can significantly deviate from their familiar idealized patterns, exhibiting irregular lobes and shallower nulls.
This irregularity is even more apparent behind the arrays, in part due to phase shifts and attenuation introduced by their enclosures, adding further complexity.
}

\edit{%
Other factors likely at play are near-field effects between the transmitter and receiver.
The far-field distance from either of our arrays is $2D^2/\lambda \approx 2.5$ meters, and the reactive/radiating near-field boundary is $0.62\sqrt{D^3/\lambda} \approx 23$ cm, where each array's largest dimension is $D \approx 11$ cm \cite{balanis}.
The separation of our arrays is $30$ cm, meaning they reside just within the radiating near-field of one another based on these rules-of-thumb.
Electromagnetic propagation in the near-field is less predictable than that in the far-field and bucks typical assumptions in array theory, most notably the plane wave assumption; recall, the beam patterns we are familiar with are typically from a far-field perspective.
In near-field settings, side lobes can grow and main lobes can widen, for instance \cite{roberts_wcm,balanis}, further complicating the interaction between transmit and receive beams.
In addition, due to these near-field effects, there is the potential that self-interference is in fact not linearly related to the transmit and receive beams as expressed in \eqref{eq:power-si}, but this requires further investigation to flesh out.
As we will see, the dominant source of self-interference actually resembles far-field interaction, raising questions about the near-field nature of the underlying self-interference channel.
To summarize, we conjecture that the combination of irregular beam patterns and near-field effects atop far-field interaction leads to seemingly random fluctuations in self-interference with slight shifts of the transmit and receive beams.
Detailed electromagnetic simulation would likely yield more concrete explanations for this log-normal behavior but would demand a significant investment of time and resources, making it a good topic for dedicated future work.
}

\comment{

Most likely, this variability over small spatial neighborhoods is tied to the fact that beam shapes in practice can significantly deviate from their familiar idealized patterns, exhibiting irregular side lobes and shallower nulls.
Presumably, self-interference couples between the back and side lobes of the transmit and receive beams.
In addition, since our arrays are mounted on separate sides of a triangular platform, the signals propagating from the transmit array elements to the receive array elements pass through the enclosures of the array modules, further distorting their beam shapes behind the arrays.

---

It is difficult to precisely describe the roots of this log-normal nature, but we can make educated speculations based on an understanding of our measurement system and wireless fundamentals. 
The most likely cause of this variability over small spatial neighborhoods is that beam shapes in practice can significantly deviate from their familiar idealized patterns.
In practice, beams exhibit irregular side lobes and shallower nulls due to electromagnetic coupling and non-isotropic and non-ideal array elements. 
In addition, since our arrays are mounted on separate sides of a triangular platform, the signals propagating from the transmit array elements to the receive array elements pass through the enclosures of the array modules. 
Consequently, each of the $256 \times 256$ paths undergo a unique phase shift and attenuation, depending on material properties and path distance.
When setting beamforming weights at the transmitter and receiver to steer their beams, this may lead to non-trivial constructive or destructive combinations due to phase and amplitude differences across array elements. 
Ultimately, this sensitive constructive/destructive combining can lead to seemingly random degrees of self-interference and, thus, may explain why self-interference follows a log-normal distribution around small spatial neighborhoods.

Another factor at play may be near-field coupling between the transmitter and receiver.
The far-field distance from either of our arrays is around $2D^2/\lambda \approx 2.5$ meters, and the reactive/radiating near-field boundary is $0.62\sqrt{D^3/\lambda} \approx 23$ cm \cite{balanis}.
The separation of our arrays is $30$ cm, meaning they reside just within the radiating near-field of one another based on these rules-of-thumb.
Electromagnetic propagation in the near-field is less predictable than that of the far-field and bucks typical assumptions in array theory, most notably the plane wave assumption.
In such settings, side lobes can grow and main lobes can widen, for instance \cite{roberts_wcm,balanis}, further complicating the interaction between transmit and receive beams.
As we will see, however, the dominant source of self-interference actually resembles far-field interaction, raising questions about the near-field nature of the true self-interference channel.
Furthermore, there is the potential that self-interference is not linearly related to the transmit and receive beams as expressed in \eqref{eq:power-si}, but this requires further investigation to flesh out.
Detailed electromagnetic simulation would likely yield more concrete explanations for this log-normal behavior but would demand a significant investment of time and resources, making it a good topic for dedicated future work.
}


\comment{
First, and perhaps most explanatory, is the \textit{near}-field interaction between our transmit and receive arrays.
The far field distance from either of our arrays is around $2.5$ meters, while their separation is only around $30$ cm, meaning they reside in the near-field of one another \cite{balanis}.
Electromagnetic propagation and behavior in the near-field is less predictable than that of the far-field and bucks typical assumptions in array signal processing, most notably the plane wave assumption.
As a result, beams deteriorate from their usual shape (e.g., \figref{fig:beam-pattern}) when observing them within the near-field: side lobes can grow and main lobes can widen, for instance \cite{roberts_wcm,balanis}.
In addition, since our arrays are mounted on separate sides of a triangular platform, the signals propagating from the transmit array elements to the receive array elements pass through the enclosures of the array modules. 
Consequently, each of the $256 \times 256$ paths may undergo a unique phase shift and attenuation, depending on material properties and the path distance, for example.
When setting beamforming vectors at the transmitter and/or receiver to steer their beams, this may lead to non-trivial constructive or destructive combinations due to minute phase and amplitude differences induced by the variety of aforementioned artifacts.
Ultimately, this sensitive constructive/destructive combining can lead to seemingly random degrees of self-interference at the output of the receive array and, thus, may explain why self-interference follows a log-normal distribution around small spatial neighborhoods.
Detailed electromagnetic simulation would likely yield more concrete explanations for this log-normal behavior but would require a significant investment of time and resources, making it a good topic for dedicated future work.

The log-normal nature of \ginr over small spatial neighborhoods---the $\nbrtwotwo$ neighborhood in particular---is the foundation for our stochastic model of \mmwave self-interference, allowing us to create a stochastic model from a single set of measurements, which themselves are deterministic.
This is in line with the goals of this work: to produce a model of \mmwave self-interference that can be used to draw realizations that spatially and statistically align with our measurements, rather than align exactly with the \ginr measured for a particular beam pair.
}

\section{Spatial and Statistical Modeling} \label{sec:modeling}

In this section, we present our model of self-interference for multi-panel full-duplex \mmwave systems.
Our method is relatively simple, has theoretical foundations, and is built on system and model parameters having real-world meaning.
Naturally, we cannot guarantee that our model---nor the measurements it is based on---will perfectly translate to other systems. 
We suspect that it can indeed be generalized to systems beyond our own, however, through appropriate parameterization, which we show in \secref{sec:remarks}.
At a high level, our model of \mmwave self-interference accomplishes two goals:
\begin{enumerate}
    \item It is stochastic, allowing researchers to draw countless realizations of self-interference, rather than a single deterministic quantity.
    \item Its realizations align spatially and in distribution with real-world self-interference.
\end{enumerate}
As mentioned in the previous section, the foundation of our model is the seemingly random nature of self-interference across small spatial neighborhoods.
Having fit a log-normal distribution to the neighborhood surrounding each beam pair $(i,j)$, we have nearly 6.5 million fitted mean $\muij$ and variance $\varij$.
In this section, we reverse-engineer a methodology to estimate these fitted parameters $\parens{\muij,\varij}$ based on transmit beam $\vf\thphtxi$ and receive beam $\vw\thphrxj$.
Then, using this estimation methodology, a realization of self-interference for general transmit beam $\vf$ and receive beam $\vw$ can be drawn as
\begin{align}
\todB{\inrreal} \sim \distgauss{\muest}{\varest} \label{eq:realize-inr}
\end{align}
where $\muest$ and $\varest$ are an estimated mean and variance based on $\vf$ and $\vw$. 


\subsection{Estimating Neighborhood Mean, $\muij$}

\begin{figure*}[!t]
    \centering
    \subfloat[Actual measurements.]{\includegraphics[width=0.45\linewidth,height=\textheight,keepaspectratio]{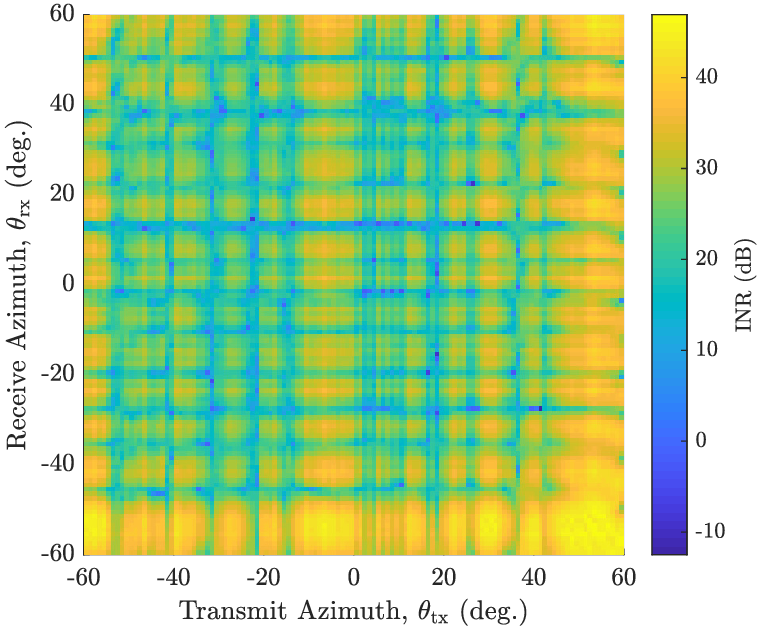}
        \label{fig:azimuth}}
    \quad
    \subfloat[Spherical-wave model \cite{spherical_2005}.]{\includegraphics[width=0.45\linewidth,height=\textheight,keepaspectratio]{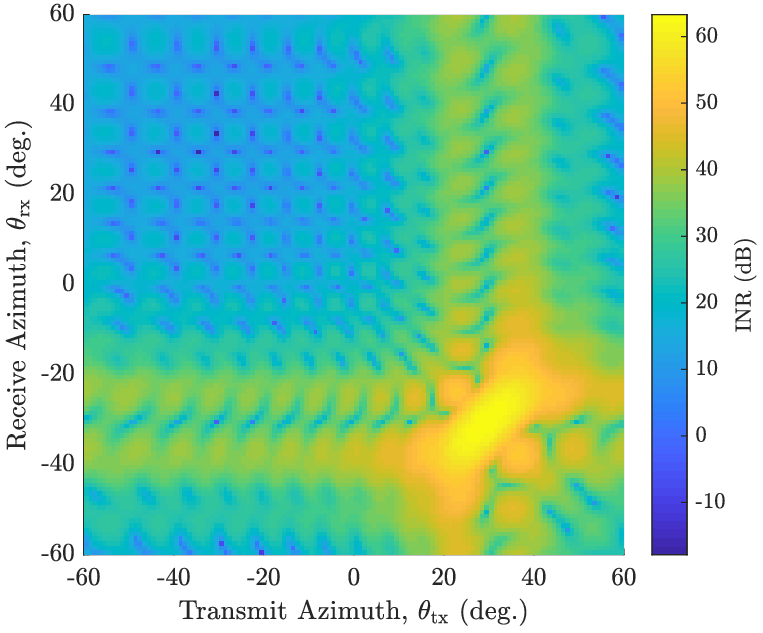}
        \label{fig:azimuth-sim}}
    \caption{\edit{(a) The measured azimuth cut of self-interference, where $\phitx = \phirx = 0^\circ$. (b) The simulated counterpart of (a) based on the commonly-used spherical-wave near-field channel model \cite{spherical_2005}. The stark difference between the two motivates the need for a new model of real-world self-interference, which we present herein.}}
    \label{fig:azimuth-meas-sim}
\end{figure*}

With nearly 6.5 million true fitted $\muij$, we seek a method to reliably estimate each based on the known transmit and receive beams used during measurement.
If successful, the same approach can then be used to estimate the mean for general transmit and receive beams.
To estimate the mean $\muij$ for a given transmit-receive pair $(i,j)$, we found that a coarse approximation of the over-the-air self-interference channel between the transmit and receive arrays can be used.
As mentioned in the introduction, the spherical-wave \mimo channel model \cite{spherical_2005}---an idealized near-field model---has been widely used in research on full-duplex \mmwave systems.
However, as pointed out in our prior work \cite{roberts_att_angular} and illustrate in \figref{fig:azimuth-meas-sim}, this spherical-wave channel model does not align with our measurements of self-interference. 
In light of this, we now propose a different channel model to coarsely describe the spatial structure of self-interference. 

\textbf{Coarse geometric modeling of the self-interference \mimo channel.} 
It is practically difficult to precisely measure the over-the-air self-interference \mimo channel (i.e., the channel matrix $\mH$ shown in \eqref{eq:power-si}) due to its high dimensionality and the fact that it is observed through the lens of the analog beamforming networks in the transmit and receive phased arrays. 
Fortunately, we have uncovered from our measurements a coarse geometric model of the self-interference channel. 
This model does not capture the significant variability observed {locally} but rather the large-scale spatial components of the self-interference channel.
We now outline this proposed geometric model, which we will use to estimate $\muij$.



\begin{figure}
    \centering
    \includegraphics[width=\linewidth,height=0.3\textheight,keepaspectratio]{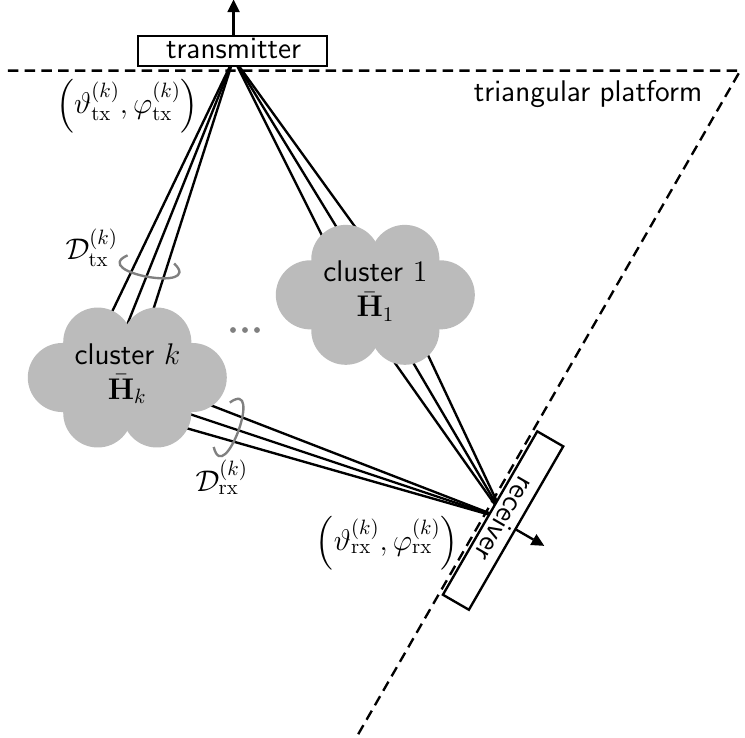}
    \caption{\edit{An illustration of coupling clusters comprising the coarse geometric model of the self-interference channel $\mHbar$ between the transmit and receive arrays. From our measurements, we observed the presence of four dominant coupling clusters. This clustered coupling is a drastic deviation from the spherical-wave model \cite{spherical_2005} that has been widely used in the literature.}}
    \label{fig:clusters}
\end{figure}

Let us begin by inspecting the azimuth cut of our \ginr measurements (i.e., $\phitx = \phirx = 0^\circ$) plotted in \figref{fig:azimuth}.
From the bright clouds in \figref{fig:azimuth}, we can readily see that there is high self-interference when $(\thetatx,\thetarx)$ is approximately $(0^\circ,-60^\circ)$, $(60^\circ,-60^\circ)$, $(60^\circ,60^\circ)$, and $(-60^\circ,-60^\circ)$.
This hints at which angular components couple high self-interference and paves the way for our proposed geometric model. 
Rather than model this coupling as single rays extending from the transmit array to the receive array, we instead model them as clouds of rays---or \textit{coupling clusters}---with some non-zero angular spread, as illustrated in \figref{fig:clusters}. 
Suppose $\numclust$ coupling clusters exist between the transmit and receive arrays, where each cluster contains the same number of rays for simplicity.
The center of the $k$-th cluster is defined by some \gls{aod} from the transmit array $\parens{\varthetatx\idx{k},\varphitx\idx{k}}$ and some \gls{aoa} at the receive array $\parens{\varthetarx\idx{k},\varphirx\idx{k}}$.
These cluster centers correspond to the bright clouds identified by visual inspection of \figref{fig:azimuth}.

We assume the angular spread of each cluster is quantified in azimuth and elevation by some $\Delta\vartheta$ and $\Delta\varphi$ and is common across clusters for simplicity.
We use $\set{D}_{\vartheta}$ and $\set{D}_{\varphi}$ defined as
\begin{align}
\set{D}_{\vartheta} &= \braces{-\Delta\vartheta,\dots,0,\dots,+\Delta\vartheta}, \qquad
\set{D}_{\varphi} = \braces{-\Delta\varphi,\dots,0,\dots,+\Delta\varphi}
\end{align}
to discretize this angular spread into individual rays with some resolution in azimuth and elevation, respectively. 
For instance, discretizing both $\set{D}_{\vartheta}$ and $\set{D}_{\varphi}$ with $1^\circ$ resolution has been sufficient for us.
We model a coupling cluster as the collection of all \gls{aod}-\gls{aoa} combinations based on the angular spread $\nbrv$.
\edit{%
In this fashion, the \glspl{aod} and \glspl{aoa} of the rays comprising the $k$-th coupling cluster can be expressed as
\begin{align}
\set{D}_{\labeltx}\idx{k} &= \parens{\varthetatx\idx{k},\varphitx\idx{k}} + \set{D}_{\vartheta} \times \set{D}_{\varphi}, \qquad
\set{D}_{\labelrx}\idx{k} = \underbrace{\parens{\varthetarx\idx{k},\varphirx\idx{k}}}_{\mathsf{cluster~center}} + \underbrace{\set{D}_{\vartheta} \times \set{D}_{\varphi}}_{\mathsf{angular~spread}}.
\end{align}
The channel matrix produced by the $k$-th cluster is then written as
\begin{align}
\mHestk &= 
\sum_{\substack{\parens{\varthetatx,\varphitx} \in \set{D}_{\labeltx}\idx{k}}} \sum_{\substack{\parens{\varthetarx,\varphirx} \in \set{D}_{\labelrx}\idx{k}}} 
\arx{\varthetarx,\varphirx} \ \atx{\varthetatx,\varphitx}\ctrans,
\end{align}
which is simply the sum of rank-$1$ matrices produced by each of the rays in the cluster; recall, $\atx{\varthetatx,\varphitx}$ and $\arx{\varthetarx,\varphirx}$ are the array responses of the transmit array and receive array for some \gls{aod} and \gls{aoa}, respectively.}
As implicitly done here, assuming all rays to have unit gain has proven sufficient for us but this could be generalized straightforwardly. 
The resulting channel matrix comprised of all $\numclust$ coupling clusters is then
\begin{align}
\mHest = \sum_{k=1}^{\numclust} \mHestk \cdot \frac{\sqrt{\Nt\cdot\Nr}}{\normfro{\sum_{k=1}^{\numclust} \mHestk}} ,
\end{align}
where the scaling is to ensure the channel matrix has fixed energy $\normfro{\mHest}^2 = \Nt \cdot \Nr$.

\begin{table}[!t]
    \small
    \centering
    \caption{Dominant cluster centers in our measurements ($\numclust = 4$).}
    \label{tab:fitting-results-clusters}
    \begin{tabular}{ccc}
        \hline
        \textbf{Cluster Index $k$} & \textbf{AoD $\parens{\varthetatx\idx{k},\varphitx\idx{k}}$} & \textbf{AoA $\parens{\varthetarx\idx{k},\varphirx\idx{k}}$} \\
        \hline
        $1$ & $(-174^\circ,0^\circ)$ & $(-122^\circ,0^\circ)$ \\
        $2$ & $(126^\circ,0^\circ)$ & $(-122^\circ,0^\circ)$ \\
        $3$ & $(-118^\circ,0^\circ)$ & $(-122^\circ,0^\circ)$ \\
        $4$ & $(126^\circ,0^\circ)$ & $(118^\circ,0^\circ)$ \\
        \hline
    \end{tabular}
\end{table}

Inspection of our measurements revealed the presence of $\numclust = 4$ dominant clusters whose centers are listed in \tabref{tab:fitting-results-clusters} and were mentioned before based on \figref{fig:azimuth}.
Notice that the azimuthal components of the \glspl{aod} and \glspl{aoa} are beyond $\pm 90^\circ$ rather than within the $[-60^\circ,60^\circ]$ that was measured, as was illustrated in \figref{fig:clusters}.
This decision is an arbitrary one since the array response of a \upa is symmetric about $\pm 90^\circ$; in other words, an azimuth of $60^\circ$ induces the same array response as an azimuth of $120^\circ$.
We made this choice to follow the intuition that self-interference is presumably coupled \textit{behind} the arrays since measurements took place in an anechoic chamber free of any significant reflectors in front of the arrays.
We have observed that the dominant coupling exists in and around the azimuth plane, meaning $\varphitx\idx{k} = \varphirx\idx{k} = 0^\circ$ for all $k = 1, \dots, \numclust$.
This can likely be attributed to the fact that our transmit and receive arrays are aligned in elevation, though in other configurations, components beyond an elevation of $0^\circ$ may play a more significant role.
From our measurements, we empirically found a good angular spread to be $\Delta\vartheta = 4^\circ$ and $\Delta\varphi = 3^\circ$. 
Note that other cluster parameters could be used for systems that see other self-interference profiles, potentially also to account for reflections off the environment.

\textbf{Using our geometric channel to estimate $\muij$.}
With our geometric channel model $\mHbar$ in hand, we now aim to estimate $\muij$ for particular transmit and receive beams.
To do so, we developed the estimator 
\begin{align}
\muijest = \todB{\frac{\eirp \cdot \Gsqest \cdot \bars{\vw\thphrxj\ctrans \mHest \vf\thphtxi}^{2\cdot\xi}}{\powernoise}}, \label{eq:compute-muij}
\end{align}
which is rooted in using $\mHest$ and the beamforming weights $\vf\thphtxi$ and $\vw\thphrxj$ to capture coupling trends as a function of transmit and receive beamforming.
The expression of \eqref{eq:compute-muij} is quite similar to and is inspired by that of \eqref{eq:power-si}--\eqref{eq:power-si-inr}. 
The scalars $\xi$ and $\Gsqest$ serve as model parameters that we will fit based on our measurements using the true $\braces{\muij}$.
Along with normalizing the energy of the channel matrix as $\normfro{\mHest}^2 = \Nt \cdot \Nr$, we also remind the reader that we normalize our beams as $\normtwo{\vf}^2 = \Nt$ and $\normtwo{\vw}^2 = \Nr$. 
These normalizations are crucial for proper scaling of $\muijest$.
Recall that \eqref{eq:compute-muij} is used to estimate the mean $\muij$ of the $(i,j)$-th neighborhood and not the actual \ginr of the $(i,j)$-th beam pair.
For this reason, $\Gsqest$ and $\mHbar$ should not be thought of as equal to $\Gsq$ and $\mH$ in \eqref{eq:power-si}, though they may closely align. 


\textbf{Fitting model parameters $\xi$ and $\Gsqest$ to measurements.}
The model parameters $\xi$ and $\Gsqest$ introduced in \eqref{eq:compute-muij} allow us to account for distributional discrepancies in $\braces{\muij}$ and the \ginr one would theoretically expect based on our coarsely-approximated channel $\mHest$.
The role of these model parameters can be better observed by expressing \eqref{eq:compute-muij} in log-form as
\begin{align}
\muijest &= \underbrace{\xi}_{\mathsf{scale}} \cdot \underbrace{\todB{\Gammaij}}_{\mathsf{shape}} + \underbrace{\todB{\Gsqest}}_{\mathsf{location}} + \underbrace{\todBm{\eirp} - \todBm{\powernoise}}_{\mathsf{system~parameters}}, \label{eq:muijest-log}
\end{align}
where 
\begin{align}
\Gammaij = \bars{\vw\thphrxj\ctrans \mHest \vf\thphtxi}^2 \label{eq:gamma-ij}.
\end{align}
From \eqref{eq:muijest-log}, we see that the shape of the distribution of $\braces{\muijest}$ is dictated by the coupling factors $\braces{\Gammaij}$, while its scale and location are controlled by $\xi$ and $\Gsqest$, respectively. 

\begin{figure*}
    \centering
    \subfloat[True fitted mean $\muij$.]{\includegraphics[width=0.475\linewidth,height=\textheight,keepaspectratio]{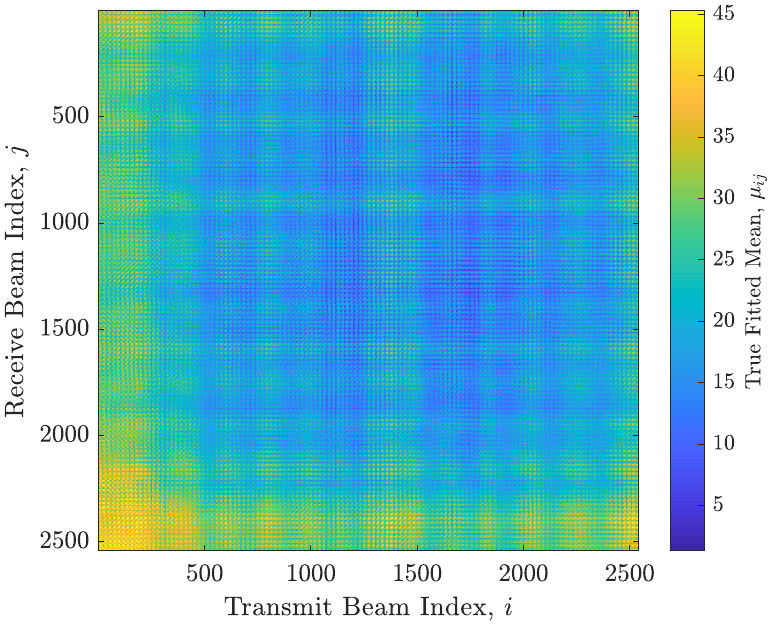}
        \label{fig:matrix-muij-a}}
    \quad
    \subfloat[Estimated mean $\muijest$.]{\includegraphics[width=0.475\linewidth,height=\textheight,keepaspectratio]{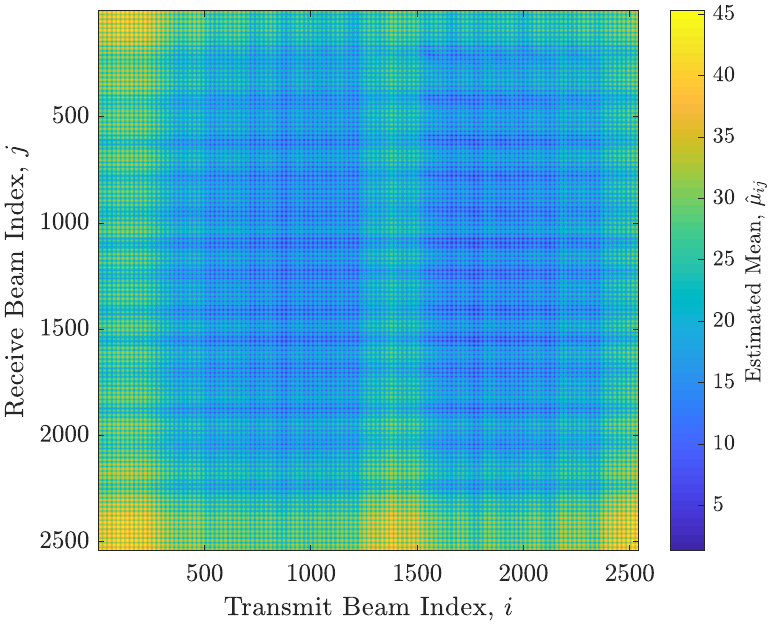}
        \label{fig:matrix-muij-b}}
    \caption{(a) The true fitted mean $\muij$ for each of the nearly 6.5 million transmit-receive beam pairs $(i,j)$. (b) The estimated counterpart of (a), where $\muijest$ for each beam pair has been estimated using \eqref{eq:muijest-log} with fitted $\xi$ and $\Gsqest$. The strong structural similarity between the two confirms that $\mHest$ captures large-scale spatial trends observed in our measurements.}
    \label{fig:matrix-muij}
\end{figure*}

In order to ensure the distribution of our estimates $\braces{\muijest}$ matches that of the true $\braces{\muij}$ in terms of scale and location, we fit $\xi$ and $\Gsqest$ as follows.
We first fit $\xi$ as
\begin{align}
\xi = \sqrt{\frac{\varop{\braces{\muij}}}{\varop{\braces{\todB{\Gamma_{ij}}}}}}
\end{align}
to align the variance of our estimates $\braces{\muijest}$ with the variance of the true $\braces{\muij}$.
Then, $\Gsqest$ is fitted to align their locations as
\begin{align}
\GsqestdB 
= \meanop{\braces{\muij}} - \meanop{\braces{\xi \cdot \todB{\Gamma_{ij}}}} - \todBm{\eirp} + \todBm{\powernoise},
\end{align}
which ensures that the mean of our estimates $\braces{\muijest}$ matches that of the true $\braces{\muij}$.
Following this fitting from our measurements and $\mHbar$, we have $\xi = 0.502$ and $\Gsqest = -129.00$ dB.




\begin{figure*}
    \centering
    \subfloat[True fitted mean $\muij$ and estimated mean $\muijest$.]{\includegraphics[width=0.475\linewidth,height=0.26\textheight,keepaspectratio]{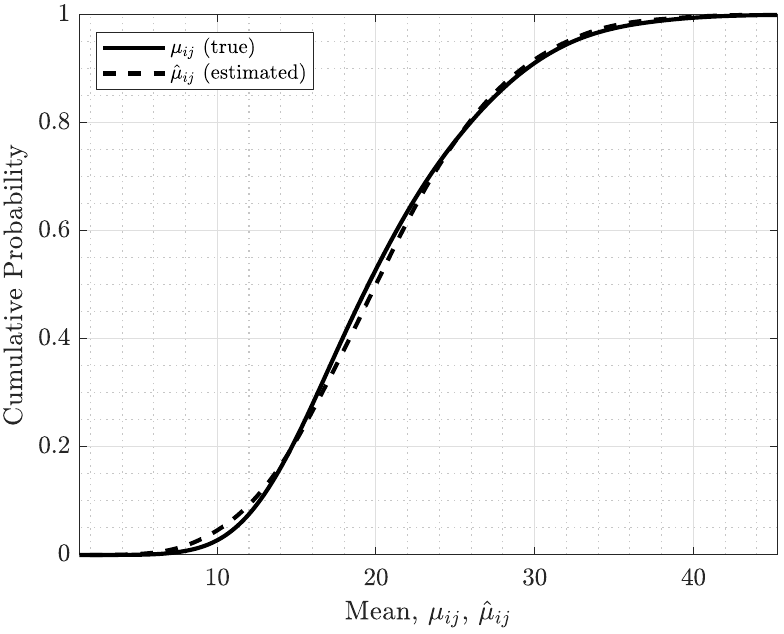}
        \label{fig:cdf-muij-a}}
    \quad
    \subfloat[Estimation error $\muij - \muijest$.]{\includegraphics[width=0.475\linewidth,height=0.26\textheight,keepaspectratio]{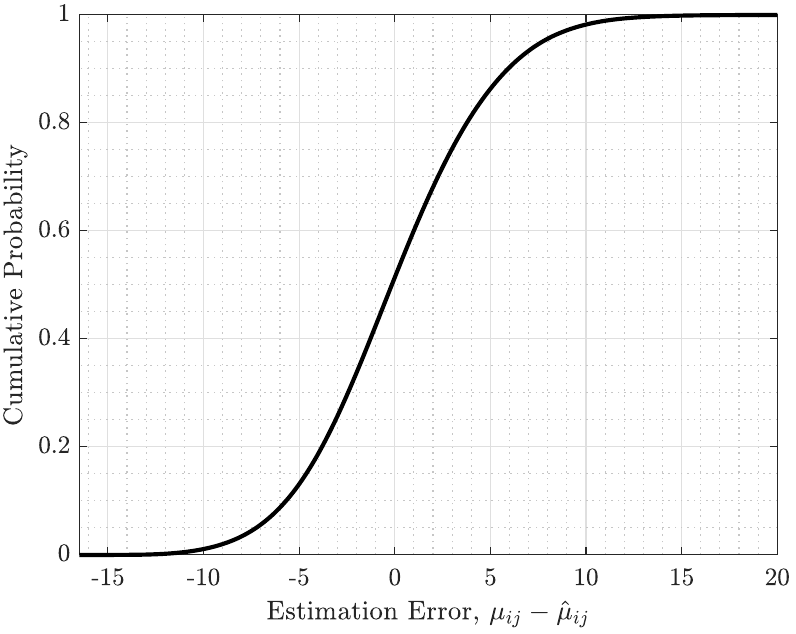}
        \label{fig:cdf-muij-b}}
    \caption{(a) The \gcdf of the true fitted mean $\muij$ and the estimated mean $\muijest$ for all nearly 6.5 million beam pairs $(i,j)$; in other words, the \gpcdf of \figref{fig:matrix-muij-a} and \figref{fig:matrix-muij-b}. (b) The \gcdf of the error between $\muij$ and $\muijest$ for all nearly 6.5 million beam pairs $(i,j)$; in other words, the \gcdf of the difference between \figref{fig:matrix-muij-a} and \figref{fig:matrix-muij-b}.}
    \label{fig:cdf-muij}
\end{figure*}

We evaluate our estimation of $\muij$ in \figref{fig:matrix-muij} and \figref{fig:cdf-muij}.
First, consider \figref{fig:matrix-muij-a} and \figref{fig:matrix-muij-b}, which depict all nearly 6.5 million $\muij$ and the corresponding estimates $\braces{\muijest}$, respectively.
Clearly, our estimates $\braces{\muijest}$ align visually with the true $\braces{\muij}$.
Our ability to closely reproduce the structure of $\braces{\muij}$ via $\mHbar$ suggests that the dominant source of self-interference stems from coupling clusters (implicitly a far-field model) as opposed to the aforementioned highly-idealized near-field model \cite{spherical_2005}.
The magnitude of these estimates $\braces{\muijest}$ follows that of the true $\braces{\muij}$, courtesy of appropriately fitted $\xi$ and $\Gsqest$.
This can be further observed in \figref{fig:cdf-muij-a}, which compares the \gpcdf of $\braces{\muij}$ and $\braces{\muijest}$ and illustrates that our estimates align closely in distribution with the ground truth.
Here, it can be thought that $\mHbar$ is responsible for aligning the \textit{shape} of the two distributions, while $\Gsqest$ and $\xi$ align the \textit{location} and \textit{scale}.
Finally, \figref{fig:cdf-muij-b} shows the empirical \gcdf of the error associated with our estimation of $\muij$.
As desired, this error is centered about $0$, and the vast majority of its density lay within $\bars{\muij-\muijest} \leq 10$.
Approximately $75$\% of our estimates are within $5$ dB and $90$\% are within $7.5$ dB.
Altogether, this evaluation confirms that our approach accurately captures the spatial and distributional characteristics of $\muij$.

\begin{figure*}
    \centering
    \subfloat[Linear estimation of $\varij$ based on $\muij$.]{\includegraphics[width=0.475\linewidth,height=\textheight,keepaspectratio]{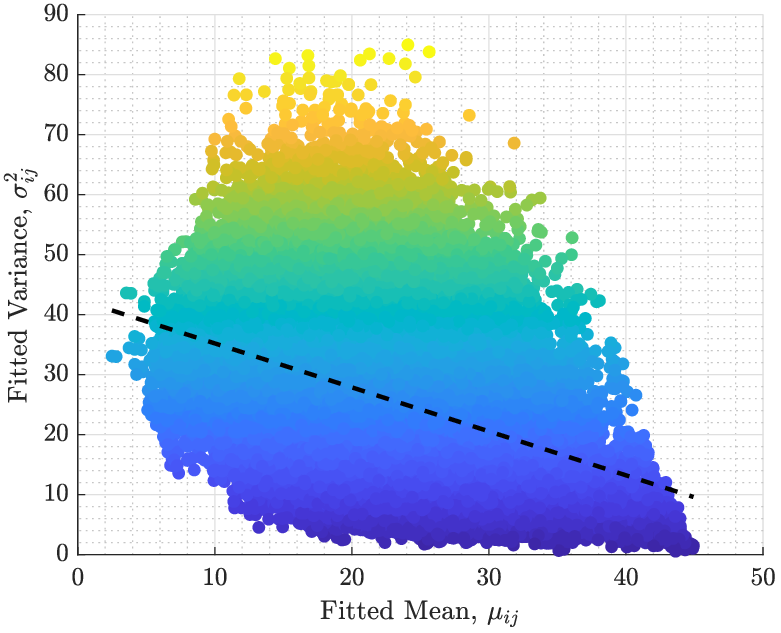}
        \label{fig:varij-a}}
    \quad
    \subfloat[Linear estimator error $\varij-\varijbar$.]{\includegraphics[width=0.475\linewidth,height=\textheight,keepaspectratio]{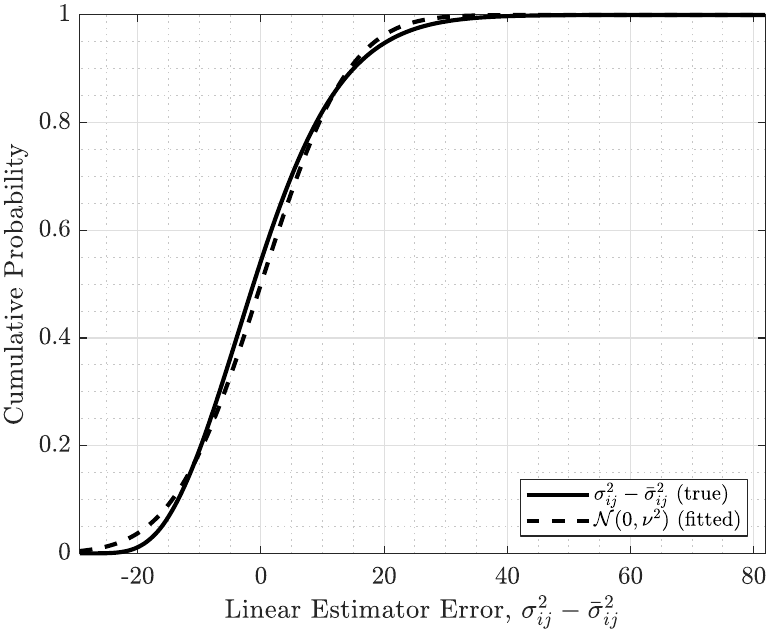}
        \label{fig:varij-b}}
    \caption{(a) Pairs of true fitted $(\muij,\varij)$ and the linear estimator (dashed line) used to produce $\varijbar = \muij \cdot \alpha + \beta$. (b) The \gcdf of the linear estimator error $\varij - \varijbar$, along with the corresponding fitted normal distribution $\distgauss{0}{\nu^2}$.}
    \label{fig:varij}
    \vspace{-0.25cm}
\end{figure*}

It is natural to ask whether $\xi$ and $\Gsqest$ hold any physical meaning.
It is difficult to state precisely, but we can interpret $\xi$ and $\Gsqest$ as accomplishing a few things.
First, most naturally, $\Gsqest$ can largely be interpreted as capturing the inverse path loss and isolation between the transmit and receive arrays (e.g., due to enclosures, free space path loss), along with amplification therein. 
The scale parameter $\xi$ plays an important role in correcting for deviations from the theoretical coupling between transmit and receive beams. 
It can account for small errors in our coarse geometric self-interference channel model $\mHest$, along with irregular beam patterns produced by practical arrays.
Practical beam patterns, for instance, typically do not exhibit the clean shapes and deep nulls we are accustomed to from array theory.
$\xi$ also may account for the fact that self-interference is potentially not linearly related to $\vf$ and $\vw$.
In addition, $\xi$ also importantly corrects for the fact that we are using $\Gammaij$ to estimate the \textit{mean} of the $(i,j)$-th neighborhood, rather than directly the \ginr of the $(i,j)$-th beam pair.
The parameters $\Gsqest$ and $\xi$---along with $\mHbar$---allow our statistical model to be extended to other systems by tailoring them appropriately and presumably contain additional underlying physical meaning beyond what has already been mentioned, even if it is not yet fully fleshed out.
Note that the estimator in \eqref{eq:compute-muij} and \eqref{eq:muijest-log} is presented for measured beam pairs but can be naturally extended to more general $\vf$ and $\vw$ by dropping $(i,j)$ indexing, using fitted $\xi$ and $\Gsqest$.


\subsection{Estimating Neighborhood Variance, $\varij$.}

\begin{figure}
    \centering
    \includegraphics[width=\linewidth,height=0.27\textheight,keepaspectratio]{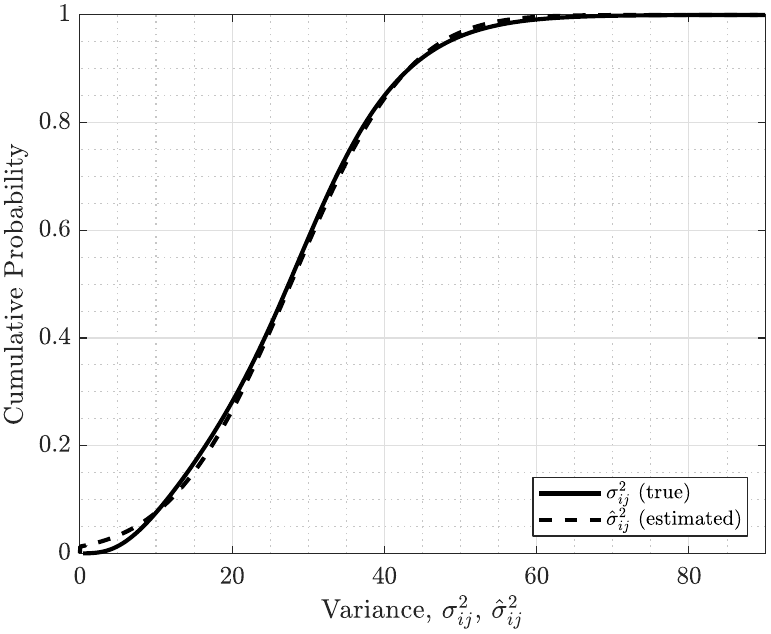}
    \caption{CDFs of the true fitted variance $\varij$ and the estimated variance $\varijest$.}
    \label{fig:cdf-varij}
\end{figure}

Having estimated the mean $\muij$, we now focus on estimating the neighborhood variance $\varij$.
Estimating $\muij$ involved model and system parameters and fairly intuitive methods with sound connections to the physical world.
The estimation of the variance $\varij$, on the other hand, proved to be much more difficult, with less pronounced dependence on the transmit and receive beams. 
Consider \figref{fig:varij-a}, which shows scattered pairs of the true fitted $\parens{\muij,\varij}$. 
We can observe that the variance $\varij$ varies significantly with the mean but that it tends to be smaller with higher $\muij$.
Conceptually, this suggests that beam pairs seeing high self-interference tend to observe less variability across their neighborhood---in line with conclusions drawn in \cite{roberts_att_angular}.
Fitting a simple linear estimator to the true $\parens{\muij,\varij}$, shown as the dashed line, allows us to make the linear estimate $\varijbar$ of the variance $\varij$ using the mean $\muij$ via
\begin{align}
\varijbar = \muij \cdot \alpha + \beta \label{eq:varijbar}
\end{align}
where $\alpha = -0.733$ and $\beta = 42.53$ are the slope and bias of the linear estimator fitted to our measurements.
While this simple estimator captures the general trend in variance as a function of the mean, it does not capture its broad and seemingly random variability observed in \figref{fig:varij-a}.
To estimate $\varijest$, we therefore include randomness from a normal distribution in conjunction with the linear estimator to increase its variability. 
Following the linear estimator, $\varijest$ is drawn via
\begin{align}
\varijest \sim \brackets{\varijbar + \distgauss{0}{\nu^2}}^{+} \label{eq:varijest}
\end{align}
where the variance of the normal distribution is fitted as $\nu^2 = 126.091$ based on the distribution of the error $\braces{\varij - \varijbar}$ illustrated in \figref{fig:varij-b}; here, $\brackets{x}^+ = \maxop{x,0}$ simply ensures $\varijest \geq 0$.

By supplementing the linear estimate $\varijbar$ with normally-distributed noise, the distribution of $\braces{\varijest}$ closely aligns with that of the true fitted $\braces{\varij}$, as illustrated in \figref{fig:cdf-varij}.
Note that this approach aligns their \textit{distributions} rather than aiming to perfectly estimate $\varijest = \varij$.
This was done deliberately due to the simple fact that there was no reliable way to true estimate $\varij$; rather, it appeared to be extremely random.
However, by aligning the distributions of $\braces{\varijest}$ and $\braces{\varij}$, our model is able to produce realizations of self-interference with spatial richness comparable to the true measurements.
This importantly allows our model to capture the variability in self-interference over small spatial neighborhoods that was observed in our measurements.



\subsection{Summary}

\begin{algorithm}[!t]
	\begin{algorithmic}[0]
        \REQUIRE $\vf$, $\vw$; $\mHest$, $\Gsqest$, $\xi$, $\alpha$, $\beta$, $\nu^2$; $\eirp$, $\powernoise$
    	\STATE \textbf{Estimate mean using \eqref{eq:compute-muij}:}
        \STATE \quad $\Gamma = \bars{\vw\ctrans \mHest \vf}^2$
        \STATE \quad $\muest = \xi \cdot \todB{\Gamma} + \todB{\Gsqest} + \todBm{\eirp} - \todBm{\powernoise}$
        \STATE \textbf{Draw variance using \eqref{eq:varijbar} and \eqref{eq:varijest}:}
        \STATE \quad $\varbar = \muest \cdot \alpha + \beta$
        \STATE \quad $\varest \sim \brackets{\varbar + \distgauss{0}{\nu^2}}^+$
        \STATE \textbf{Draw self-interference:}
        \STATE \quad $[\inrreal]_{\mathrm{dB}} \sim \distgauss{\muest}{\varest}$
        \STATE \textbf{Bound \ginr to prevent extreme outliers (optional):}
        \STATE \quad $\inrreal \leftarrow \minop{\maxop{{\inrreal},{\inrmin}},{\inrmax}}$
        \STATE \textbf{Calculate self-interference power (if needed):}
        \STATE \quad $\todBm{\powersi} = \todBm{\powernoise} + [\inrreal]_{\mathrm{dB}}$
	\end{algorithmic}
	\caption{A statistical realization of self-interference.}
	\label{alg:algorithm-realization}
\end{algorithm}

    \edit{A summary of our statistical model of self-interference is shown in \algref{alg:algorithm-realization}. 
    For particular transmit and receive beamforming weights $\vf$ and $\vw$, the mean $\muest$ is computed using model parameters $\mHest$, $\Gsqest$, and $\xi$ and system parameters $\eirp$ and $\powernoise$.
    Then, the variance $\varest$ can be realized using $\muest$ and model parameters $\alpha$ and $\beta$, along with normally-distributed noise having variance $\nu^2$.
    Recall, these system and model parameters can be unique to a particular full-duplex \mmwave platform; they may be fitted directly from measurements or referenced from future wireless standards or the literature.
    A realization of self-interference can be drawn from the normal distribution $\distgauss{\muest}{\varest}$, which can be bounded if desired to prevent extreme outliers.}
\section{Generalization and Applications of our Statistical Model} \label{sec:remarks}
In the previous section, we detailed the construction of our model based on a set of nearly 6.5 million measurements.
In this section, we highlight how this model can generalize to other systems by collecting nearly 13 million additional measurements from two different system configurations.
Then, we highlight applications of our proposed model to overview how it can be used in full-duplex research and development.

\subsection{Fitting Our Model to Other Configurations}

Naturally, we cannot guarantee how well our statistical model will translate to other systems, carrier frequencies, and environments, but it does have the potential to via custom system and model parameters.
Our model readily accommodates general system \gls{eirp} and noise power. 
\edit{Model parameters can be uniquely tailored to individual platforms by collecting measurements and subsequently fitting them using the methods described in \secref{sec:modeling} or by using parameters published in future literature or wireless standards.}
Other systems will presumably see unique spatial profiles of self-interference, which can be accounted via $\mHest$ through a different number of clusters, different cluster \glspl{aod}/\glspl{aoa}, and a different angular spread.
Note that our model can directly accommodate arbitrary array geometries and sizes by constructing $\mHest$ with the appropriate array responses.
However, it is not clear how different array geometries may affect our model and its fitted parameters.
For instance, perhaps the small-scale variability observed over small spatial neighborhoods decreases with fewer array elements. 
Investigating this could make for interesting future work.


\begin{table*}[!t]
    \small
    \centering
    \caption{System parameters and fitted model parameters from measurements.}
    \label{tab:fitting-results}
    \begin{tabular}{ccccc}
        \hline
        \textbf{Parameter(s)} & \textbf{Symbol(s)} & \textbf{Default} & \textbf{Vertical} & \textbf{Tapered} \\
        \hline
        EIRP & \textbf{$\eirp$} & $60$ dBm & $60$ dBm & $54$ dBm \\
        Noise Power & \textbf{$\powernoise$} & $-68$ dBm & $-68$ dBm & $-68$ dBm \\
        \hline
        {Cluster Centers} of $\mHbar$ & ${\parens{\varthetatx\idx{k},\varphitx\idx{k}}, \parens{\varthetarx\idx{k},\varphirx\idx{k}}}$ & See \tabref{tab:fitting-results-clusters} & See \tabref{tab:fitting-results-clusters} & See \tabref{tab:fitting-results-clusters} \\
        Angular Spread of $\mHbar$ & \textbf{$(\Delta\vartheta,\Delta\varphi)$} & $(4,3)$ & $(4,3)$ & $(4,3)$ \\
        Location Parameter of $\muest$ & \textbf{$\Gsqest$} & $-129.00$ dB & $-141.58$ dB  & $-144.58$ dB \\
        Scale Parameter of $\muest$ & \textbf{$\xi$} & $0.502$ & $0.527$ & $0.498$ \\
        \hline
        Estimator Parameters of $\varest$ & \textbf{$(\alpha,\beta)$} & $(-0.733,42.53)$ & $(-0.588,29.71)$ & $(-0.822,25.42)$ \\
        Variance Parameter of $\varest$ & $\nu^2$ & $126.091$ & $75.794$ & $110.391$ \\
        \hline
    \end{tabular}
\end{table*}


To more concretely confirm that our model can indeed generalize beyond the single set of measurements it was constructed on, we repeated our nearly 6.5 million measurements of self-interference in two additional system configurations. 
We fit our model of self-interference to each set of measurements and tabulated their fitted parameters in \tabref{tab:fitting-results}:
\begin{itemize}
    \item The \textit{default} column corresponds to the aforementioned setup and nearly 6.5 million measurements used extensively herein and presented in \cite{roberts_att_angular}.
    \item The \textit{vertical} column corresponds to a vertically-polarized version of \textit{default}, where we repeated the measurements with both arrays physically rotated $90^\circ$.
    \item The \textit{tapered} column corresponds to repeating the measurements of \textit{default}, except with transmit and receive beams having tapered side lobes. 
\end{itemize}
When fitting our model to each of these, we found that the spatial profile of self-interference was common across all three configurations.
In other words, the same coarse model of the self-interference channel $\mHest$ was observed in all three cases. 
This exciting finding reinforces that the coupling between the arrays occurs in the form of clusters of rays and that this coupling is a function of the relative array geometry and of the mounting infrastructure. 

We can observe that $\xi$ is approximately $0.5$ across all three configurations, suggesting it too is tied to the physical setup and/or the arrays themselves.
Relative to the default configuration, $\Gsqest$ decreases in the vertical and tapered configurations.
In the vertical configuration, this is perhaps courtesy of extra isolation offered by the array enclosures and elements when rotated sideways.
In the tapered configuration, this decrease in $\Gsqest$ is perhaps due to the extra isolation offered by reduced side lobe and back lobe levels. 
Connections between the other model parameters across the three measurements is less obvious, and fleshing out how system factors impact all model parameters is an attractive topic for future work.

\comment{
In addition, we can observe that $\Gsqest$ is nearly identical in the default and tapered cases, further confirming that $\Gsqest$ captures the inherent isolation between the arrays.
Put simply, since these two were collected with the same physical setup and orientation, it is expected that they have the same inherent isolation.
The $12$ dB increase in $\Gsqest$ in the vertically polarized orientation is perhaps courtesy of extra isolation offered by the array enclosures and elements when rotated sideways.
This suggests that $\xi$ is presumably tied to beam shape and how such may deteriorate in practice due to nonidealities.
Connections between the other model parameters across the three measurements is less obvious, and fleshing out how system factors impact all model parameters is an attractive topic for future work.
}

While not explicitly shown due to space constraints, the additional two measurement sets yielded fitted models that offered similar results as those presented for the default setup in \secref{sec:modeling}.
This confirms that our model can indeed be generalized to other systems through unique parameterization. 
Furthermore, the fact that the fitted model parameters of all three measurement sets are of the same order---and even closely align in some cases---strengthens the belief that our model and its parameters have strong underpinnings to the real world, even if they are not yet fully understood.
\edit{In general, however, the viability of and methodology for extending our model to systems beyond our own requires contributions and measurements from the research community at large. 
Our hope is that the model presented herein is a first step toward even more rigorous modeling of self-interference in full-duplex \mmwave systems, with the end goal being a practical model of the self-interference channel matrix $\mH$, if possible.}

%

\comment{
\begin{remark}[Generalizing our model to other systems and settings]
Naturally, we cannot guarantee how well our statistical model will translate to other systems and environments but have made an attempt to give it the potential to via system and model parameters.
For instance, a system with increased isolation between its transmit and receive arrays may decrease $\Gsqest$ to capture such in our model.
Likewise, a system transmitting at a different \gls{eirp} or having a different noise floor, can account for through $\eirp$ and $\powernoise$.
In addition, the approximated self-interference channel matrix $\mHest$ may be varied by tweaking the number of clusters, cluster locations, and cluster angular spread.
For instance, if one wants to consider other array geometries (i.e., other than a triangular platform), perhaps other cluster locations would be more appropriate.
Note that our model can accommodate arbitrary array geometries and sizes by constructing $\mHest$ with the appropriate array responses.
However, it is not clear how different array geometries may affect our model and its fitted parameters.
For instance, perhaps the small-scale variability observed over small spatial neighborhoods (i.e., $\sigma^2$) decreases with fewer array elements.
This makes for interesting future work.

Unfortunately, we cannot comment on the potential levels of self-interference a system would expect with arbitrary transmit and receive beams $\vf$ and $\vw$.
Our measurements were taken using highly directional conjugate beams as defined in \eqref{eq:beams-f-w}.
As a result, we are only confident in our model when considering beams of this form, though we do expect similarly shaped beams to also work with our model.
In general, beams $\vf$ and $\vw$ do not have to necessarily lay on our grid of measured beams. 
They can potentially be within our $1^\circ$ resolution or beyond our measured spatial profiles in \eqref{eq:spatial-profile}, but we cannot be sure of this with verifying such through measurement.
In general, the viability of and methodology for extending our model to systems beyond our own requires plenty of dedicated future work from the research community. 
Ultimately, our hope is that the model presented herein is a first step toward even more rigorous and universal modeling of self-interference in full-duplex \mmwave systems, with the end goal being a practical model of the self-interference channel matrix $\mH$, if possible.



\end{remark}
}

\subsection{Applications of Our Model} \label{subsec:apps}
Our model allows engineers to draw statistical realizations of self-interference that a practical full-duplex \mmwave system would incur when using particular transmit and receive beams $\vf$ and $\vw$.
A full-duplex \mmwave system employing codebook-based beamforming, for example, will see unique levels of self-interference between each transmit-receive beam pair---our model can be used to realize these levels of self-interference.
In such a case, transmit-receive beam pairs offering low self-interference will be more desirable in a full-duplex sense.
This simple fact raises interesting questions about the potential for user selection to facilitate full-duplex operation, which can be further investigated using our statistical model.
Also, note that beams $\vf$ and $\vw$ do not necessarily have to lay on our grid of measured beams. 
They can potentially be within our $1^\circ$ resolution or beyond our measured spatial profiles defined in \eqref{eq:spatial-profile}, but we cannot be sure of the reliability of such with verifying it through measurement.
\edit{Our model also readily accommodates the case where transmit and receive beams steer in the same direction---i.e., $\thphtx = \thphrx$---such as when communicating with another full-duplex device.}



Another important application of our model is its role in creating and evaluating beamforming-based solutions for full-duplex \mmwave systems.
While we have evaluated our model using conjugate beams and beams with tapered side lobes, we cannot confidently expect our model to produce accurate realizations of self-interference with \textit{arbitrary} transmit and receive beams $\vf$ and $\vw$; this would require extremely precise modeling of the self-interference channel $\mH$. 
As a result, we can only be confident in our model when considering highly directional beams, especially the conjugate beams we have used in \eqref{eq:cbf-f-w}. 
Therefore, our model should not be used directly to evaluate beamforming-based full-duplex solutions that tailor transmit and receive beams based on knowledge of the true self-interference channel $\mH$.

Nonetheless, our model is suitable for evaluating some beamforming-based solutions, especially those that rely purely on beam steering to reduce self-interference.
\steer \cite{roberts_steer}, for instance, is a beam refinement methodology that leverages the variability of self-interference over small spatial neighborhoods to significantly reduce self-interference without compromising downlink and uplink beamforming gain.
Our statistical model can be used to evaluate \steer, as we will show in the next section, and can facilitate the creation and evaluation of other beamforming-based solutions for full-duplex beyond \steer.
Since our model is stochastic one, it can be drawn from indefinitely to conduct research over a broad variety of realizations of self-interference---perhaps network-wide---which can shed light on potential edge-cases and lower-tail performance.
Finally, our model's realizations of self-interference can also be used to drive requirements on the degree of additional cancellation required by full-duplex \mmwave systems employing analog or digital \sic.





\comment{
\begin{align}
\todB{\inrijest} \sim \distgauss{\muijest}{\varijest}
\end{align}

\begin{align}
\distgaussijest &\triangleq \distgauss{\muijest}{\varijest} 
\end{align}

As illustrated in \figref{fig:clusters}, suppose $\numclust$ coupling clusters exist between the transmit and receive arrays, where each cluster contains $\numrays$ rays for simplicity.
The center of the $i$-th cluster is defined by some \gls{aod} from the transmit array $\parens{\varthetatx\idx{i},\varphitx\idx{i}}$ and \gls{aoa} at the receive array $\parens{\varthetarx\idx{i},\varphirx\idx{i}}$.
The set of coupling cluster centers is simply
\begin{align}
\braces{\parens{\underbrace{\varthetatx\idx{i},\varphitx\idx{i}}_{\mathsf{AoD}},\underbrace{\varthetarx\idx{i},\varphirx\idx{i}}_{\mathsf{AoA}}} : i = 1, \dots, \numclust}.
\end{align}

To fit the scale of the distribution of the estimates $\braces{\muijest}$ with that of the true $\braces{\muij}$, we fit $\xi$ based on our collected measurements as
\begin{align}
\xi = \sqrt{\frac{\varop{\braces{\muij}}}{\varop{\braces{\todB{\Gamma_{ij}}}}}}
\end{align}
which ensures that 
\begin{align}
\varop{\braces{\muijest}} = \varop{\braces{\muij}}.
\end{align}

To fit $G^{2}$ used for estimating $\muij$, we use
\begin{align}
\todB{G^{2}} 
= \meanop{\braces{\muij}} - \meanop{\braces{\xi \cdot \todB{\Gamma_{ij}} + \todBm{\powertx} - \todBm{\powernoise}}}
\end{align}
which will ensure that the mean of our estimates $\braces{\muijest}$ matches that of the true $\braces{\muij}$.
\begin{align}
\meanop{\braces{\muijest}} = \meanop{\braces{\muij}}
\end{align}
With $\mHest$ sufficiently modeled and $\xi$ and $G^{2}$ fitted, the shape, scale, and location of the distribution of $\braces{\muijest}$ are well aligned with that of $\braces{\muij}$, which we illustrate next.

}

\section{Evaluating our Statistical Model} \label{sec:evaluation-ii}
In this section, we evaluate our model of self-interference by comparing realizations drawn from our model against those observed in measurements.
We aim to show that our model is capable of capturing the large-scale spatial trends and the small-scale spatial variability seen in our measurements.
As part of this, we use our model to evaluate an existing full-duplex solution and compare the results to an evaluation using actual measurements.






\subsection{Comparing Our Statistical Model to Measurements}
Ideally, values of self-interference drawn from our model should align spatially and in distribution with what was measured. 
To evaluate such quantitatively, consider the following.
Let $\sKtx$ be a random subset of $K$ elements from $\braces{1,\dots,\Ntx}$ without replacement.
In other words, $\sKtx$ contains $K$ random transmit indices.
Let $\sKrx$ be defined analogously as a set of $K$ random receive indices.
\begin{align}
\sKtx &= \mathrm{randk}\parens{\braces{1,\dots,\Ntx},K}, \qquad
\sKrx = \mathrm{randk}\parens{\braces{1,\dots,\Nrx},K}
\end{align}
Then, let $\sM$ be the set of measured \ginr values for each transmit-receive beam pair whose indices are elements of $\sKtx$ and $\sKrx$, defined as
\begin{align}
\sM  = \braces{\todB{\inr_{mn}} : m \in \sKtx, \ n \in \sKrx},
\end{align}
where we employ the shorthand $\inr_{mn} \triangleq \inr\parens{\thetatx\idx{m},\phitx\idx{m},\thetarx\idx{n},\phirx\idx{n}}$.
We can draw a statistical realization of $\sM$ by forming $\sMreal$ as
\begin{align}
\sMreal = \braces{\todB{\inrmnreal} \sim \distgauss{\mumnest}{\varmnest} : m \in \sKtx, \ n \in \sKrx},
\end{align}
where $\inrmnreal$ is drawn from our statistical model. 
By comparing $\sM$ and $\sMreal$, we can gain insight into how well our model can realize self-interference that is statistically and spatially similar to our measurements.


\begin{figure*}
    \centering
    \subfloat[Actual measurements.]{\includegraphics[width=0.32\linewidth,height=0.28\textheight,keepaspectratio]{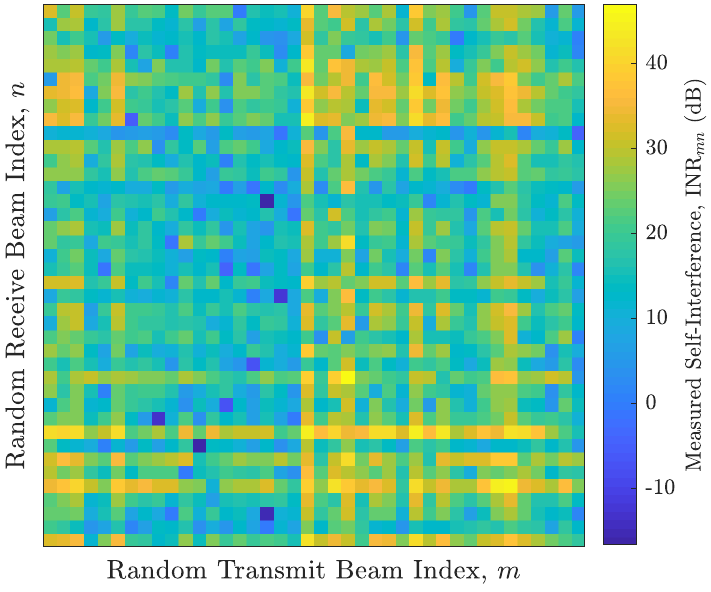}
        \label{fig:matrix-a}}
    \hfill
    \subfloat[Proposed model.]{\includegraphics[width=0.32\linewidth,height=0.28\textheight,keepaspectratio]{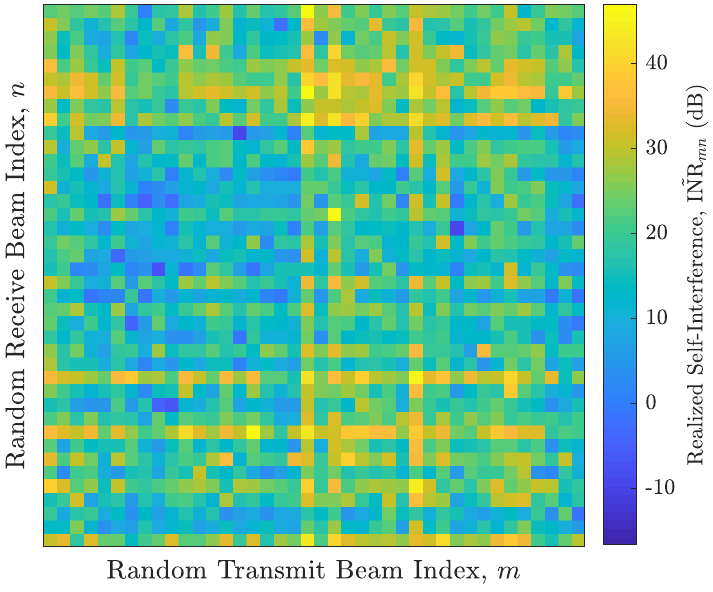}
        \label{fig:matrix-b}}
    \hfill
    \subfloat[Spherical-wave model \cite{spherical_2005}.]{\includegraphics[width=0.32\linewidth,height=0.28\textheight,keepaspectratio]{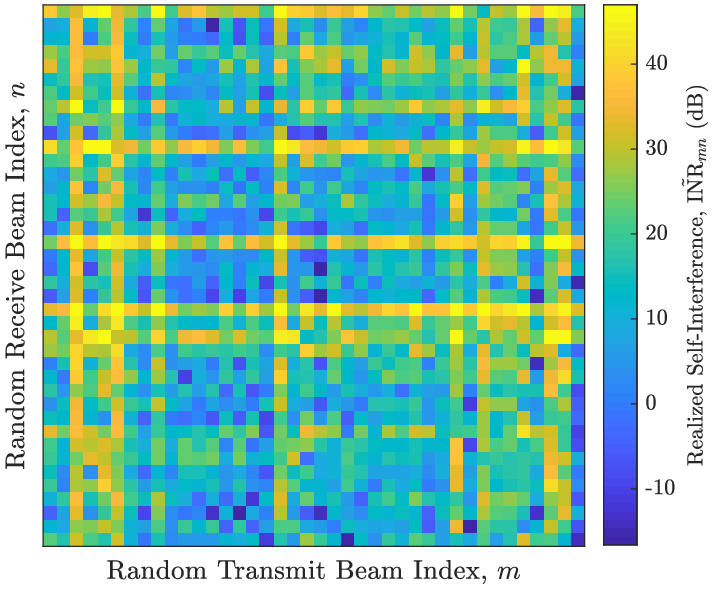}
        \label{fig:matrix-c}}
    \caption{\edit{(a) Measured self-interference $\sM$ for $K=40$ random transmit beams and $K=40$ random receive beams. (b) A realized counterpart $\sMreal$ of (a) using our statistical model. (c) A realized counterpart of (a) using the spherical-wave model \cite{spherical_2005}. The structural similarity between (a) and (b) illustrates that our model captures the dominant spatial components of self-interference in our measurements, whereas (c) does not.}}
    \label{fig:matrix}
    \vspace{-0.5cm}
\end{figure*}


Consider \figref{fig:matrix}, which compares measurements $\sM$ and realizations $\sMreal$ of self-interference for $K = 40$ random transmit beams and receive beams; \edit{also shown is self-interference one would expect based on the spherical-wave model \cite{spherical_2005} that has been widely used in the literature.\footnote{\edit{To model self-interference under the spherical-wave channel \cite{spherical_2005}, we constructed $\mH$ based on the geometry of our phased array platform and used \eqref{eq:power-si}, where $G^2$ was fitted to align in median with our measurements of self-interference.}}}
Notice the color scale is common across all three. 
Visually, one can observe the strong structural similarity between \figref{fig:matrix-a} and \figref{fig:matrix-b}.
The transmit beams (columns) that tend to inflict higher self-interference in our measurements also tend to when using our model to realize self-interference.
As observed in our measurements, these transmit beams do not yield high self-interference universally.
Rather, there are select receive beams that couple low self-interference, even if the majority couple high self-interference.
Likewise can be said about the receive beams (rows) that tend to couple higher degrees of self-interference.
Similar conclusions are drawn when considering beams that tend to yield low self-interference, though these are less pronounced in general.
Comparing \figref{fig:matrix-a} and \figref{fig:matrix-b} importantly demonstrates that our model is capable of realizing self-interference that coincides spatially with our measurements.
\edit{\figref{fig:matrix-c}, on the other hand, illustrates that the spherical-wave model \cite{spherical_2005} does not at all align with our measurements of real-world self-interference---observed also in \figref{fig:azimuth-meas-sim}.}

\begin{figure*}
    \centering
    \subfloat[\edit{Measured and realized distributions.}]{\includegraphics[width=\linewidth,height=0.27\textheight,keepaspectratio]{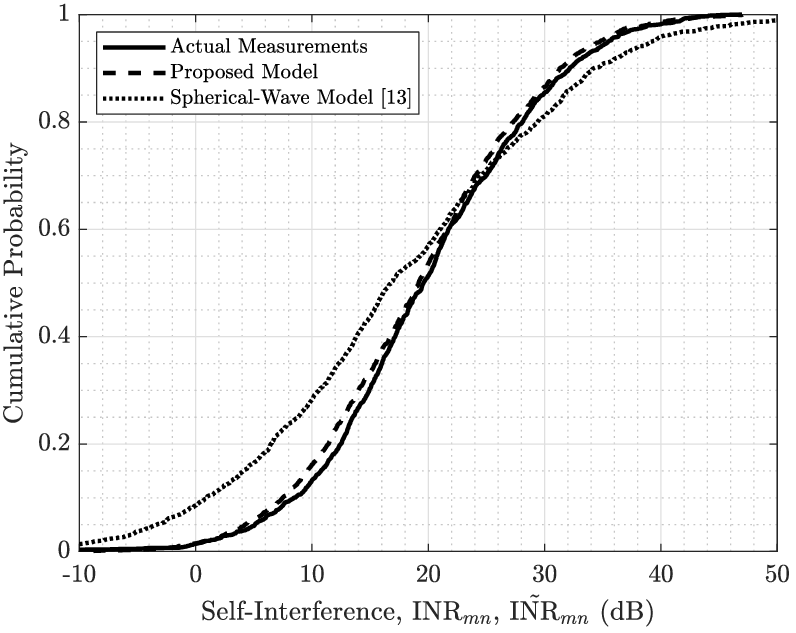}
        \label{fig:cdf-matrix}}
    \hfill
    \subfloat[Assessing convergence in distribution.]{\includegraphics[width=\linewidth,height=0.27\textheight,keepaspectratio]{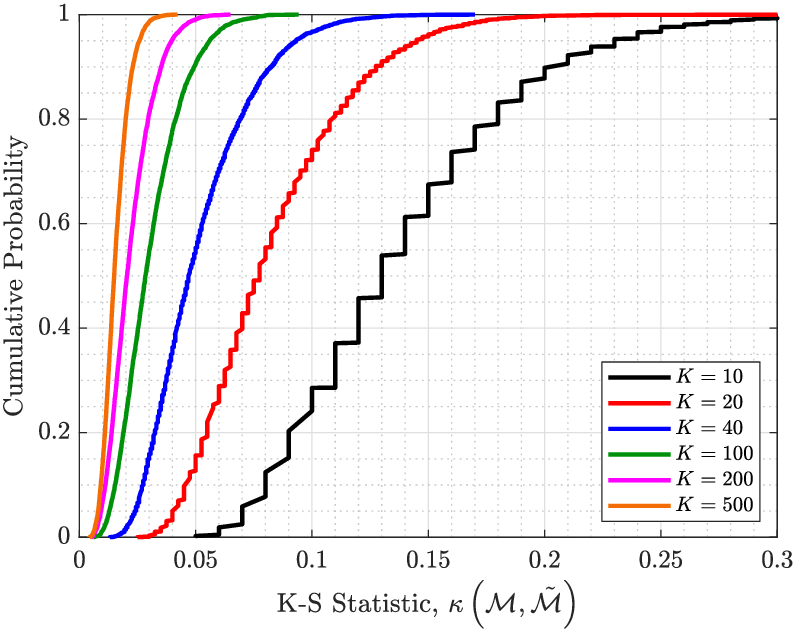}       \label{fig:cdf-ks}}
    \caption{(a) The \gpcdf of the measured and realized self-interference in \figref{fig:matrix-a} and \figref{fig:matrix-b}, aligned closely in distribution. Also shown is the \gcdf of \figref{fig:matrix-c} based on the spherical-wave model \cite{spherical_2005}, which aligns neither structurally nor in distribution with measurements. (b) The \gpcdf of the \gls{ks} statistic $\kappa\parens{\sM,\sMreal}$ between measured and realized self-interference, where each \gcdf is taken across $5000$ random $\parens{\sM,\sMreal}$ for each $K$. As $K$ increases, the measured and realized self-interference more closely align in distribution.}
    \label{fig:cdf}
    \vspace{-0.25cm}
\end{figure*}

The \gpcdf of the three figures in \figref{fig:matrix} are depicted in \figref{fig:cdf-matrix}, which shows that our model produces self-interference that is statistically similar to that of measurements.
While our model may not \textit{exactly} replicate what has been measured (by design), it importantly does yield \textit{distributions} of self-interference in line with measurements while also capturing dominant spatial characteristics.
\edit{The spherical-wave model \cite{spherical_2005}, however, does not align closely in distribution with actual measurements, exhibiting heavier upper and lower tails.}



Now, consider \figref{fig:cdf-ks}, where we extend this comparison of measurements and our proposed model over many realizations and for various $K$.
For a given $K$, we draw $5000$ random measurements $\sM$ and realizations $\sMreal$ with our model, and for each, we compute the K-S statistic $\kappa\parens{\sM,\sMreal}$.
The K-S statistic here is the difference in density between the \gpcdf of $\sM$ and $\sMreal$ and desirably is less than $0.1$.
\figref{fig:cdf-ks} depicts the resulting \gcdf of the K-S statistic over all $5000$ random $\parens{\sM,\sMreal}$.
With $K=10$, the K-S statistic lay below $0.2$ for $90$\% of the random realizations.
As $K$ increases, our model produces realizations of self-interference $\sMreal$ that \edit{more closely align in distribution} to actual measurements $\sM$.
With $K = 40$, for instance, over $96$\% of the time, our model produces realizations of $\sMreal$ whose distribution differs from measurements $\sM$ by at most $0.1$ in density.
This suggests that our model can be reliably used to realize self-interference that aligns statistically with our measurements, even with a fairly modest number of samples.


\edit{The comparison of our statistical model to measurements in \figref{fig:matrix} and \figref{fig:cdf} is particularly applicable in the context of codebook-based full-duplex \mmwave systems, as mentioned in \subsecref{subsec:apps}.
A system employing transmit and receive codebooks, each with $K$ beams for instance, will see a unique degree of self-interference between each of the $K^2$ transmit-receive beam pairs. 
With our model, engineers can draw realizations of self-interference between each beam pair, producing a matrix of self-interference that aligns spatially and statistically with our measurements, especially if the number of beams is at least around $K=20$.}

\subsection{Using Our Model to Design and Evaluate Solutions}
Recent work \cite{roberts_steer} proposes a beamforming-based solution for full-duplex \mmwave systems called \steer. 
\steer is a measurement-driven beam refinement methodology for full-duplex \mmwave systems that can be applied atop conventional codebook-based beam selection. 
Suppose initial transmit and receive beam selections (i.e., steering directions) have been made at a full-duplex \mmwave transceiver to serve some downlink user and uplink user.
Surrounding these initial transmit and receive steering directions, \steer constructs spatial neighborhoods based on some resolution and size. 
Each transmit-receive steering combination within the spatial neighborhood is measured until a beam pair offering sufficiently low self-interference is found, minimizing the distance the selected beam pair is from the initial beam selection.
As such, the success of \steer as a full-duplex solution relies heavily on the variability of self-interference over small spatial neighborhoods.
Therefore, to confirm our model can be used to accurately evaluate \steer, it must produce realizations of self-interference that offer variability over small spatial neighborhoods that is of a similar degree as that seen in reality. 

To investigate if our model can be used to evaluate full-duplex solutions, we replicate the evaluation process of \steer described in \cite{roberts_steer}, with minor changes stated as follows; we refer the reader to \cite{roberts_steer} for details not explicitly stated herein.
We first use measurements of self-interference to evaluate \steer and then use realizations of self-interference from our statistical model.
If the two evaluations align, this will provide further confidence in the relevance and practicality of our model.
When running \steer, we use a spatial resolution of $\nbrd = \nbroneone$ and a neighborhood size of $\nbr = \nbrtwotwo$, along with a self-interference target of $\inr \leq 0$ dB.
On the downlink and uplink, we perform initial beam selection using a codebook of beams distributed in azimuth from $-56^\circ$ to $56^\circ$ with $8^\circ$ spacing and in elevation from $-8^\circ$ to $8^\circ$ with $8^\circ$ spacing, for a total of 45 beams.
We use exhaustive beam search to maximize downlink and uplink \gsnr when conducting conventional beam selection.

\begin{figure*}
    \centering
    \subfloat[\gcdf of self-interference.]{\includegraphics[width=\linewidth,height=0.27\textheight,keepaspectratio]{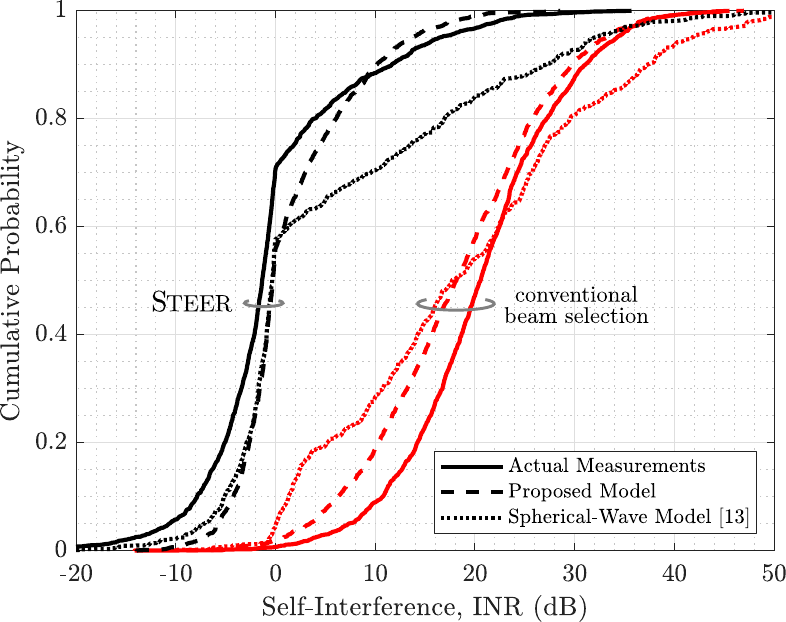}
        \label{fig:eval-a}}
    \hfill
    \subfloat[Sum spectral efficiency.]{\includegraphics[width=\linewidth,height=0.27\textheight,keepaspectratio]{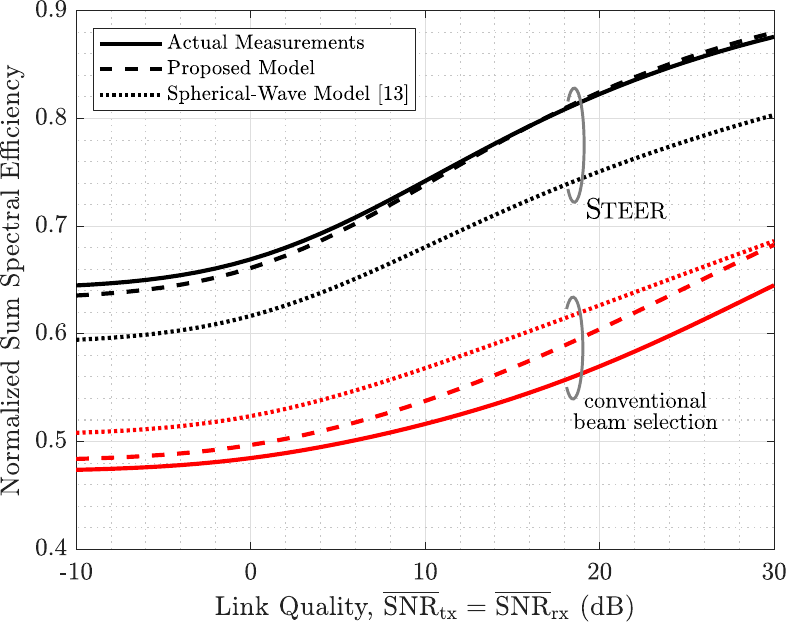}
        \label{fig:eval-b}}
    \caption{\edit{Evaluating a beamforming-based full-duplex solution \steer \cite{roberts_steer} using actual measurements, using our proposed model, and using the spherical-wave model \cite{spherical_2005}. The (a) \gcdf of self-interference and (b) normalized sum spectral efficiency when using \steer versus conventional beam selection. Our proposed model more closely aligns with measurements and therefore serves as a more reliable evaluation tool.}} 
    \label{fig:eval}
    \vspace{-0.65cm}
\end{figure*}

The results of our evaluations of \steer are shown in \figref{fig:eval}.
First, consider \figref{fig:eval-a}, which depicts the distribution of self-interference with and without \steer, when using actual measurements for evaluation (solid lines), when using our statistical model (dashed lines), \edit{and when using the spherical-wave model \cite{spherical_2005} (dotted lines).}
Without \steer (red lines), the distribution of self-interference across transmit-receive beam pairs following conventional beam selection lay largely above $0$ dB for the most part, with most beam pairs yielding self-interference levels prohibitively high for full-duplex.
The similarity between the solid and dashed red lines confirms that our model can realize levels of self-interference comparable to reality (i.e., akin to \figref{fig:matrix} and \figref{fig:cdf-matrix}). 
With \steer (black lines), lower levels of self-interference can be reached much more reliably: a significant fraction of the time, self-interference is near or below the noise floor.
Albeit not perfect, the solid and dashed black lines closely coincide, which importantly shows that our statistical model produces small-scale variability comparable to that seen in real systems.
\edit{The spherical-wave model \cite{spherical_2005} falls short in both cases, yielding distributions that do not as closely follow the actual measurements.}

We extend this evaluation a step further by inspecting the resultant sum spectral efficiency in \figref{fig:eval-b} as a function of downlink and uplink quality.
Again, evaluation with our statistical model closely aligns with that using actual measurements, differing in sum spectral efficiency only marginally, while the spherical-wave model \cite{spherical_2005} differs more substantially.
From \figref{fig:eval-a} and \figref{fig:eval-b}, we do note, however, that our model exhibits slightly more optimistic levels of self-interference with conventional beam selection and slightly more pessimistic levels with \steer, when compared to measurements.
Nonetheless, it is clear that our proposed model captures the large-scale trends in self-interference and the small-scale variability observed in practice, confirming that it can indeed be used to develop and evaluate solutions for full-duplex \mmwave systems\edit{---more closely aligning with actual measurements than the spherical-wave model \cite{spherical_2005}.}



\section{Conclusion and Topics for Future Work} \label{sec:conclusion}

In this paper, we presented the first measurement-backed model of self-interference in multi-panel full-duplex \mmwave systems.
The proposed model can produce realizations of self-interference that are statistically and spatially aligned with our collected measurements.
We have effectively reduced the presumably quite complex self-interference channel to a model parameterized by a few values, offering convenience and intuition on the underlying nature of the channel.
With appropriate parameterization, our model can translate to systems beyond our own, and we encourage others to fit, evaluate, and publish model parameters to the research community.
Importantly, we uncovered a coarse self-interference channel model that suggests that energy propagates from the transmit array to the receive array along clusters of rays in a far-field fashion---a drastic deviation from the spherical-wave model commonly used in the literature thus far.
We showed that our model can produce realizations of self-interference that exhibit the large-scale and small-scale trends observed in real systems, making it a useful tool in the design and evaluation of full-duplex solutions and systems---especially for those without access to phased array platforms.


This work motivates and enables a wide variety of future research.
\edit{First, independent work confirming that our model is applicable to a variety of full-duplex systems at various carrier frequencies and in a multitude of environments would be an essential next step, along with reporting fitted model parameters for such.}
From this, it would be useful to establish connections between system parameters and model parameters (e.g., how array size impacts $\alpha$, $\beta$, and $\nu^2$).
Naturally, we hope our model can serve as useful tool in the advancement of full-duplex \mmwave systems by allowing researchers to conduct statistical performance analyses and design full-duplex solutions.
Training machine learning models to estimate or produce realizations of self-interference would also be interesting future work.
A lofty and important goal for the research community is to develop a model of the self-interference channel (i.e., $\mH$) that is backed by measurements; it is our hope that the findings in this paper---especially the coarse channel $\mHest$---can be useful leads in this pursuit.







\section*{Acknowledgments}

I.~P.~Roberts is supported by the National Science Foundation Graduate Research Fellowship Program under Grant No.~DGE-1610403. 
Any opinions, findings, and conclusions or recommendations expressed in this material are those of the authors and do not necessarily reflect the views of the National Science Foundation.


\bibliographystyle{bibtex/IEEEtran}
\bibliography{bibtex/IEEEabrv,refs}

\end{document}